\documentclass[12pt]{article}
\pdfoutput=1
\usepackage{xtex}
\usepackage{graphicx}
\usepackage{subfigure} 

\newcommand{\beq}{\begin{equation}}
\newcommand{\eeq}{\end{equation}}
\newcommand{\beqa}{\begin{eqnarray}}
\newcommand{\eeqa}{\end{eqnarray}}

\newcommand{\half}{\ensuremath{\frac{1}{2}}}

\def\be{\begin{equation}}
\def\ee{\end{equation}}
\def\ba{\begin{eqnarray}}
\def\ea{\end{eqnarray}}

\def\tg{\widetilde{g}}
\def\tV{\widetilde{V}}
\def\tP{\widetilde{P}}
\def\half{\frac{1}{2}}
\def\uinf{u_\infty}
\def\lie{{\cal L}}

\begin{document}

\begin{titlepage}

\setcounter{page}{1} \baselineskip=15.5pt \thispagestyle{empty}


\begin{center}
{\LARGE Shock waves in strongly coupled plasmas II}
\end{center}
\vspace{\bigskipamount}

\begin{center}
{\large Sergei Khlebnikov, 
 Martin Kruczenski 
 and Georgios Michalogiorgakis
}
\end{center}

\begin{center}
\textit{Department of Physics, Purdue University \\ 
525 Northwestern Avenue, West Lafayette, IN 47907} 
\end{center} 
\vskip .01 in
\begin{center}
{\tt skhleb@physics.purdue.edu \quad markru@purdue.edu \quad gmichalo@purdue.edu } 
\end{center}
\vfil

 \noindent 
In a recent paper we have analyzed the AdS/CFT duals to shock waves propagating in the ${\mathcal N}=4$ plasma. Here we study further
properties of the system. In the gravity description we consider the properties of the dual black holes,
showing in particular that they are stationary black holes with expanding horizons.  This is possible because the horizon is not 
compact; in the fluid, this corresponds to the 
situation when entropy is being produced and carried away to infinity.
We also consider shocks in dimensionalities $d$ other than four and find that, for plasmas whose duals are given by asymptotically AdS 
spaces, the exponential 
tail of the shock on the supersonic side shrinks as $\gamma^{-\frac{2}{d}}$ as the velocity approaches the speed of light
(the Lorentz factor $\gamma$ goes to infinity). This generalizes the behavior $\gamma^{-\half}$ we have found previously 
for $d=4$.
Finally, we consider corrugations of the shock front and show that the shock is stable under such perturbations. There are,
however, long lived 
modes, excitations of which describe generation of sound by the shock wave, the energy for this 
being provided by the incoming fluid. 
 
\vfil

\end{titlepage}


\tableofcontents

\clearpage

\section{Introduction and summary}\label{INTRO}

 A useful tool for studying gauge theories at strong coupling is provided by the AdS/CFT correspondence \cite{Maldacena:1997re,Gubser:1998bc,Witten:1998qj,Aharony:1999ti}.  Lately much effort has been concentrated on studying the dynamics of strongly coupled plasmas.  Several reviews can be found in \cite{Son:2007vk,Shuryak:2008eq,Schafer:2009dj,Gubser:2009md,Gubser:2009sn,Hubeny:2010ry}.  In conformal plasmas at strong coupling the hydrodynamic description is valid up to length scales of order of the inverse temperature.  At shorter scales, hydrodynamics is not valid and one must study the microscopics of the theory.  For theories with $AdS$ duals, a way beyond the hydrodynamic description is to study the dual gravity theory, an approximation which does not break until the much smaller string length scale.  Within that context we wish to study a specific phenomenon, shock waves in conformal plasmas. 

 Shock waves play an important role in the dynamic of fluids in the supersonic regime. In recent years,
a particular fluid, the quark gluon plasma, has been studied experimentally and evidence has been found for
creation of a Mach cone by a moving heavy quark. This QCD plasma is difficult to study theoretically
because it is in the regime of strong coupling. One intriguing possibility to gain some insight into its properties 
is to use the AdS/CFT
correspondence, which maps the physics of a closely related plasma to the physics of black holes in $AdS$ space. The plasma in question is a conformal plasma that appears when the $\mathcal{N}=4$ SYM is heated to a finite temperature. Although not the same as the QCD plasma, it is strongly coupled and may have similar properties, at least at the qualitative level. 

Shock waves can appear during the initial stages of creation and thermalization of the Quark Gluon plasma.  They create entropy and might be responsible for the whole initial entropy creation; in certain models, as for example in  \cite{Landau:1953gs}, the latter determines the number of produced particles.  Shocks can also appear when a hard parton moves with a supersonic velocity.  In this case, one expects that they are closely related to the parton's Mach cone.  Several aspects of this process have been studied within AdS/CFT \cite{Friess:2006fk,Gubser:2007xz,Gubser:2007ga,Chesler:2007an,Chesler:2007sv}.  In the heavy ion collisions literature the existence of shock waves and their consequences have been studied both theoretically and experimentally \cite{Scheid:1974zz,Baumgardt:1975qv,Gutbrod:1989wd,Gutbrod:1989gh,Adams:2003kv,Adare:2008qa,Wang:2004kfa,Adams:2005ph,Adler:2005ee,Ulery:2005cc,Ajitanand:2006is,Adare:2008cqb,Stoecker:2004qu,Ruppert:2005uz,Koch:2005sx,CasalderreySolana:2004qm,Bouras:2009nn,Bouras:2010nt}.    

In a recent paper \cite{Khlebnikov:2010yt}, we have observed that, since shock waves probe the microscopic of the theory,
they are in the regime where, according to AdS/CFT, the dual gravity description becomes important. More precisely, for a perturbation of wavelength much larger that $1/T$, where $T$ is the temperature, we can just use hydrodynamics but for shorter wavelengths the appropriate microscopic description is the gravity dual.  In fact, the correspondence allows us to work both ways: 
when considering weak shocks, which can be resolved in hydrodynamics, we have learned about the existence of dual 
black holes with solitary waves propagating on their horizons, while when considering strong shocks we have been able to
use gravity to compute the penetration depth of the exponential tails on both sides of the shock and thus learn something 
about the fluid. 

 In this paper we continue our investigation of these issues. We study in more detail 
the gravity solutions corresponding to weak shocks, especially, in regard to the properties of their horizon and the fact 
that they create entropy even when they are stationary.  For strong shocks, we study the properties of the exponential tails 
in dimensionalities other than four and show that, just as in $d=4$, for shocks propagating close to the speed of light 
($\gamma \to \infty$), the scaling of the penetration depth on the supersonic side is determined by the near-boundary
region of the dual geometry and, in this sense, is universal for all fluids with asymptotically $AdS$ duals.
We also explore a region close to the horizon where there are potential divergences. Finally, we investigate the 
question of corrugation stability of the shock and show that the shock can produce sound, which travels 
in a definite direction that depends only on the velocity of the shock.

\section{Shock waves in ideal and first order hydrodynamics}\label{HYDRO}
\subsection{Shock waves in ideal hydrodynamics}\label{IDEALHYDRO} 

In this section we examine shock waves in ideal hydrodynamics in arbitrary dimensions.  The stress energy tensor for an ideal conformal fluid in $(d-1)$ space and one time dimension can be written as 
\eqn{IDEALDTENSOR}{
T^{(0)}_{\m\n} = p (g_{\m\n} +d u_{\m}u_{\n}) \;,\quad p= \frac{(4 \pi T) ^{d}}{16 \pi G_{N}^{[d+1]} d^{d}}\;,
}
where $p$ is the pressure, $T$ the temperature, $u_{\m}$ the $d$-velocity field, and $G_{N}^{[d+1]}$ is a constant that in holographic computations is naturally identified with Newton's constant in the dual $d+1$ dimensional gravity model.  We follow the notation of \cite{Bhattacharyya:2008mz} which arises from the holographic computation of \eno{IDEALDTENSOR}.  For reference we give the value of $G_{N}^{[d+1]}$ in three interesting cases.  The most well studied AdS/CFT duality is between $d=4$, $\mathcal{N}=4$ $SU(N)$ super Yang-Mills theory and type $IIB$ string theory on $AdS_{5} \times S^{5}$.  In this case $G_{N}^{[5]}= \frac{\pi}{2 N^2}$.  For $d=3$, $\mathcal{N}=6$ superconformal $U(N)_{k}\times U(N)_{-k}$ Chern-Simons theory is believed to be dual to M-theory on $AdS_{5}\times S^{5}$ \cite{Aharony:2008ug}, with $G_{N}^{[4]}=\frac{1}{N^2} \sqrt{\frac{9 \l}{8}}$ where $\l =\frac{N}{k}$ is the 't Hooft coupling.  For the $d=6$ superconformal theory living on $N$ $M5$ branes we have that $G_{N}^{[7]}=\frac{3 \pi^2}{16 N^3}$.  It should be noted that in the following whenever we do not explicitly give the dependence on $G_{N}^{[d+1]}$ we have normalized the stress energy tensor by $16 \pi G_{N}^{[d+1]}$.       

  As in \cite{Khlebnikov:2010yt} we study a shock moving in the $x$ direction in the frame in which the shock front is static.  The two components of the stress tensor that are constant are $T_{tx}$ and $T_{xx}$.  In the present case, these are given by 
\eqn{CONSERVEDCOMP}{
T^{(0)}_{tx} = p d  \frac{v}{1-v^2}\;,\quad T^{(0)}_{xx} = p(1+ d \frac{v^2}{1-v^2})\;.
}
Labelling the fields on the supersonic side with subscript 1 and those on the subsonic side with subscript 2,
we find that the fields are related by
\eqn{GOTCONN}{
v_{1}v_{2} = \frac{1}{d-1}\;, \quad \frac{p_{2}}{p_{1}}= \frac{(d-1)^2v_{1}^2-1}{(d-1)(1-v_{1}^2)}\;, \quad \frac{T_{2}}{T_{1}}=\left(\frac{(d-1)^2v_{1}^2-1}{(d-1)(1-v_{1}^2)}\right)^{1/d}\;.
}
\begin{figure}
 \begin{center}
\includegraphics[height=4cm,width=6cm]{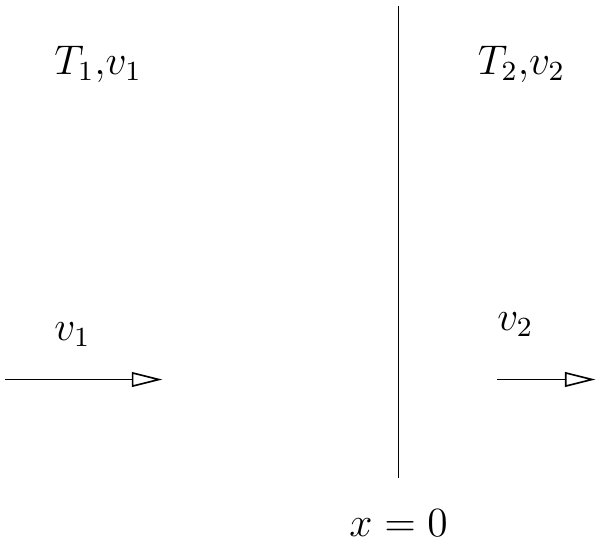}
 \end{center}
 \caption[]{Sketch of a shock wave in the rest frame of the interface.  For ideal hydrodynamics there is a discontinuity at $x=0$.  Including the higher order terms in the expression of the stress energy tensor resolves the discontinuity.  The conventions of the paper are that the fluid moves to the right. The left hand side of the flow is supersonic and the right hand side is subsonic,
the choice prescribed by the second law of thermodynamics.}
\label{ZeroShockFig}
 \end{figure}
For $d=4$ these agree with the expressions derived in \cite{Khlebnikov:2010yt}.  It should be noted here that we study shocks for $d>2$.  The case $d=2$ is special since the speed of sound coincides with the speed of light.  There are no supersonic flows for conformal fluids in two spacetime dimensions.   This is slightly disappointing: as noted in \cite{Bhattacharyya:2008mz} conformal fluid dynamics in two dimensions is trivial \footnote{Since left and right moving waves do not interact in two dimensions there is no local equilibration process.}.  There are no non-zero Weyl covariant tensors that can be constructed from the velocity and temperature fields. 

The shocks are the only mechanism that generates entropy in ideal hydrodynamics. The difference in the entropy
flux between the two regions is given by
\eqn{ENTROPYDIFF}{
s_{1,x}-s_{2,x} = 4\pi \left( \frac{4 \pi T_{1}}{d} \right)^{d-1} \left( \frac{ v_{1} }{ \sqrt{1-v_{1}^2 }}-\frac{\left((d-1)^2 v_{1}^2-1\right)^{\frac{d-2}{2d}} }{\left((d-1)(1-v_{1}^2)\right)^{\frac{d-1}{d}}}\right) \;.
}

\subsection{Shock waves in first order hydrodynamics}\label{FIRSTORDER}

When we include effects of viscosity, the profile of the shock is no longer discontinuous.  The temperature and velocity fields are varying continuously.  The first correction to the ideal stress energy tensor \eno{IDEALDTENSOR} is given by 
\cite{LL,Son:2007vk}
\eqn{STRESSFIRST}{
T^{(1)}_{\m\n} = - 2 \h \s_{\m\n}\;, \quad \h= \frac{s}{4\pi}=\frac{(4\pi T)^{d-1}}{16\pi G_{N}^{[d+1]} d^{d-1}}\;,
}
\eqn{GOTSIGMA}{
P_{\m\n} = \h_{\m\n} +u_{\m}u_{\n}\;, \quad \s^{\m\n} = P^{\m \a} P^{\n \b} \p_{( \a}u_{\b)} - \frac{1}{d-1} P^{\m\n} 
\p_{\a}u^{\a}\;.
}
The equations of motion are given by $\partial _{\m} T^{\m\n}=0$, where the stress energy tensor is now 
the sum of \eno{IDEALDTENSOR} and \eno{STRESSFIRST}. The nontrivial equations are those 
for the $T_{tx}$ and $T_{xx}$ components. To obtain these components, we first compute
\eqn{GOTSIGMAS}{
\s_{tx} = \frac{d-2}{d-1}u_{x}(x)\sqrt{1+u_{x}(x)^2} u_{x}'(x) \;, 
\quad \s_{xx} =\frac{d-2}{d-1}u_{x} '(x)\left(1+u_{x}(x)^2\right)\;. 
}
Then, the components of the stress energy tensor are
\eqn{GOTTTXFIRST}{
T_{tx} = \left(\frac{4 \pi T}{d}\right)^{d-1} u_{x}(x)\sqrt{1+u_{x}(x)}\left(4 \pi T - 2 \frac{d-2}{d-1} u_{x} '(x) \right)\;,
}
\eqn{GOTTXXFIRST}{
T_{xx}=\left(\frac{4 \pi T}{d}\right)^{d-1} \left(  \frac{4 \pi T}{d} (1+d u_{x}(x)^2)-2 \frac{d-2}{d-1}(1+u_{x}(x)^2)u_{x} '(x) \right)\;.
}

We expect the first order hydrodynamics to be a good approximation for perturbations whose characteristic wave numbers 
in the rest frame of the fluid are small compared to the temperature:
\eqn{GRADCOND}{
|q'| = \gamma |q| \ll T \;.
}
It turns out that for shock waves this condition is satisfied only when
the velocities on both sides of the shock are close to the speed of sound.  We then 
search for a solution for the $x$ component of the $d$-velocity
field in the form
\eqn{VELDEF}{
u_{x}(x) = \sqrt{\frac{1}{d-2}} + u_{\infty} \d u (x) \;,
}
where $u_{\infty} = u_x(-\infty)$ is a small parameter. It is convenient to introduce, instead of $x$, a rescaled variable 
\eqn{XIDEF}{
\xi = \frac{2 \pi T (d-2) u_{\infty}}{d-1} x \;.
}
To the leading nontrivial order in $u_{\infty}$, the 
equation of motion for $\d u(\xi)$ becomes
\eqn{GOTEOMVEL}{
\d u '(\xi) =  \d u(\xi)^2-1  \;,
}
which has the following solution:
\eqn{GOTSOLVEL}{
\d u(\xi) = \tanh(-\xi)\;.
}
The temperature perturbation can then be obtained by noting that the stress energy tensor satisfies 
the ``Landau gauge'' condition
\eqn{LANDAUGAUGE}{
T_{\m\n}u^{\m} = -(d-1)\frac{(4\pi T)^d}{d^{d}}u_{\n}\;,
}
or more explicitly
\eqn{GOTTFIRST}{
T=\frac{d}{4\pi (d-1)^{1/d}}\left( T^{xx}-T^{tx} \frac{u_{t}}{u_{x}} \right)^{1/d}\;.
}
Asymptotically, the behavior of the solution is exponentially decaying: 
\eqn{GOTASYMPVEL}{
u_\pm(x) \sim \sqrt{\frac{1}{d-2}} \mp  u_{\infty} + A_\pm e^{i \frac{q}{\pi T} x} \;,}
 with 
\eqn{GOTQSOL}{
\frac{iq}{\pi T} =4 \sqrt{\frac{d-1}{d-2}}v_{\infty}= 4 \frac{d-2}{d-1}u_{\infty} \;,
}
where 
\eqn{VUREL}{
u_{\infty}= \left(\frac{d-1}{d-2}\right)^{3/2}v_{\infty}\;.
}

Another way to use the first-order hydrodynamics is to perform linearized analysis around a fluid moving uniformly 
at constant temperature. Formally, this can be done for any fluid velocity, but as we will see in a moment it
is meaningful only for velocities close to the speed of sound. Assuming the following form for the velocity
and temperature fields 
\eqn{ASYMPVAR}{
u(x) = \frac{v}{\sqrt{1-v^2}} + \d u  e^{\frac{iqx}{\pi T}}\;, \quad T(x) = T + \d T e^{\frac{iqx}{\pi T}}\;
}
and linearizing the equations of motion, one obtains a $ 2 \times 2$ linear
system for $\d u$ and $\d T$. 
For this system to have a nontrivial solution the determinant must be zero and this determines $q$ to be 
\eqn{GOTQLIN}{
\frac{i q}{\pi T} = \frac{2\sqrt{1-v^2}}{(d-2)v}\left(v^2 (d-1)-1\right)\;.
}
For $v$ close to the speed of sound, $|q|$ is small, and \eno{GOTQLIN} agrees with the result \eno{GOTQSOL} obtained 
from the solution to the nonlinear problem. For other values of $v$, the small-gradient condition \eno{GRADCOND}
breaks down and the first-order theory is inapplicable. Note in particular that for ultrarrelativistic
speeds ($v\to 1$) $q$ as given by \eno{GOTQLIN} tends to zero in disagreement with the result that we obtain from the analysis of the black hole quasinormal modes in Section \ref{STRONG}. 

The entropy current in the first order hydrodynamics is given by 
$s^{\m} = \frac{1}{16 \pi G_{N}^{[d+1]}}\frac{4\pi (4\pi T)^d}{d^d} u^{\m}$ and satisfies 
\eqn{FIRSTENERGY}{
\partial_{\m}s^{\m} = \frac{2\h}{T} \s_{\m\n}\s^{\m\n}=\frac{2\h}{T(x)} \frac{d-2}{d-1} u_{x}'(x)^2 \;.
}
Thus, the entropy production in the first order theory is symmetric around the position of the shock front 
($x=0$).\footnote{In the next (second) order hydrodynamics, there is an ambiguity in the definition and divergence of the entropy current.  For a relevant discussion one can see \cite{Bhattacharyya:2008xc}. }

\section{Linear analysis for strong shocks}\label{STRONG}

Strong shocks are characterized by large gradients of the velocity and temperature.  As such, they cannot be analyzed within hydrodynamics.  Since the gauge/gravity correspondence is not limited to large wavelengths, it can provide information about strong shocks. In particular, we can carry out linearized analysis near a uniform flow directly in the dual gravitational theory.

Consider then a uniformly boosted black hole in $d+1$ dimensions. The metric is
\eqn{BOOSTEDBH}{
ds^2= \frac{dz^2}{z^2 \left(1-(\frac{z}{z_{h}})^d\right)} +\frac{z^{d-2}}{z_{h}^{d}} \left(\cosh \b dt -\sinh \b dx \right)^2 + \frac{-dt^2+d \vec{x} ^2}{z^2}\;.
}
For convenience we have set $L$, the radius of AdS equal to $1$.  The parameter $\b$ is related to the velocity of the boost $v$ via
\eqn{BDEF}{
v= \tanh \b\;.
}
The temperature of the dual plasma can be computed from the location of the horizon $z_{h}$ and we find that
\eqn{ZHDEF}{
T = \frac{d }{4 \pi z_{h}}\;.
}
Via the standard holographic dictionary \cite{Gubser:1998bc,Witten:1998qj},
the metric \eno{BOOSTEDBH}  gives the stress energy tensor \eno{IDEALDTENSOR} of an ideal fluid moving with velocity $v$.  
The normalization of \eno{IDEALDTENSOR} has been chosen so that it naturally arises from the gravity computation.  
We are interested in perturbations in the sound channel, which are given by
\eqn{PERTURBBH}{
ds^{2}_{pert} = \frac{e^{iqx}}{z^2}\left( H_{00}dt^2 + H_{11} dx^2 + 2 H_{01} dt dx + H d \vec{x}_{d-2}^2\right)\;.
}    
Interestingly, in the case of the sound mode the following linear combination decouples and can be studied separately: 
\eqn{GOTSOUND}{
Z(z) = H_{00}(z) + g(z) H(z)\;, \quad g(z) = 1+\frac{d-2}{2}\frac{z^d}{1-v^2}\;.
}
We are considering here a perturbation with no time dependence: $\oo =0$, that is we are asking what is the penetration length for a static perturbation around a boosted black hole. 
If we think of that perturbation as being due to a shock wave present somewhere
in the fluid, the penetration length can be interpreted as the width of the exponential tail of the shock.

The linearized equation of motion for $Z(z)$ can be derived from Einstein's equations as explained in appendix \eno{DERIVATION} and is
\eqn{EOMZ}{
Z''(z) + P(z) Z'(z)+ Q(z) Z(z) =0\;,
}
where
\eqn{GOTP}{
P(z)=\frac{-(2d-1)z^{2d}+z^d \left(3d-1-2\frac{d-1}{d-2}(1-v^2)\right)-2\frac{(d-1)^2}{d-2}(1-v^2) }{z(z^d-1)\left(z^d -2\frac{d-1}{d-2}(1-v^2)\right)} \;,
}
\eqn{GOTQ}{
Q(z)=\frac{d^2 z^{2d-2}}{(z^d-1)\left(z^d-2\frac{d-1}{d-2}(1-v^2) \right)}+q^2 \frac{z^d+1-v^2}{(z^d-1)^2(1-v^2)} \;.
}
In \eno{GOTSOUND}-\eno{GOTQ} we have set $z_{h}=1$.  The full dependence on the location of the horizon $z_{h}$ can be recovered by the rescaling 
\eqn{RESCALE}{
z \rightarrow \frac{z}{z_{h}}\;, \quad q\rightarrow q z_{h}\;.
}

\subsubsection{Behavior close to the pole} \label{sec:close_to_pole}

The equations of motion for the sound mode have a pole at 
\eqn{GOTFUNNYPOINT}{
z_{f} =  \left(2\frac{d-1}{d-2}(1-v^2)\right)^{1/d}\;.
}
The pole coincides with the location of the horizon when 
\eqn{HORPOLE}{
z_{f}=1 \; \Rightarrow v_{f}= \sqrt{\frac{d}{2(d-1)}} \;.
}
This velocity is always greater than the speed of sound.  For velocities greater than $v_{f}$ the pole lies between the boundary and the horizon, the physical region for which we wish to solve \eno{EOMZ}. Therefore it is important to understand the behavior of $Z(z)$ close to the pole.  Expanding \eno{EOMZ} around $z_{f}$ we find that there are two solutions    
\eqn{ZPOLE}{
Z(z) \sim (z-z_{f})^{s}\;,\quad  s=0\;, \quad s=3\;.
}
Fuch's theorem \cite{Arfken} guarantees that the larger root $s=3$ corresponds to a regular solution.  The second solution can in general have the form: 
\eqn{ZFUNSERIES}{
Z(z) \sim \sum_{n=0} c_{n} (z-z_{f})^{n} + \sum_{n=0} \tilde{c}_{n+3} (z-z_{f})^{n+3} \log (z-z_{f})\;,
} 
 that is there can be logarithmic terms starting with $(z-z_{f})^3 \log (z-z_{f})$.  
Whether such logarithms are in fact present depends on the precise values of the coeffcients in the expansions of $P(z)$ and $Q(z)$:
\eqn{PSERZF}{
P(z) = \frac{p_{-1}}{z-z_{f}} + \sum_{n=0}^{\infty} p_{n} (z-z_{f})^n \;,
}
\eqn{QSERZF}{
Q(z) = \frac{q_{-1}}{z-z_{f}} + \sum_{n=0}^{\infty} q_{n} (z-z_{f})^n\;.
}
Substituting \eno{ZFUNSERIES} into the equation \eno{EOMZ}, we obtain a recursion relation for the coefficients $c_n$, $\tilde{c}_n$.
Specifically, cancellation of the $(z-z_f)^{-1}$ poles gives
 \eqn{GOTC1}{
 c_{1} = c_{0}\frac{z_{f}^{d-1}}{2(z_{f}^d-1)} \; , 
 }  
cancellation of the $(z-z_f)^0$ terms gives
 \eqn{GOTC2}{
 c_{2} = c_{0}  \frac{2q^2+z_{f}^{d-2}(d-1)(d-2)}{4(z_{f}^d-1)(d-2)}d \;,
 }
while cancellation of the $(z-z_f)^1$ terms determines $\tilde{c}_3$. For $\tilde{c}_3$ to vanish, 
the coefficients of \eno{PSERZF},\eno{QSERZF} have to satisfy 
\eqn{GOTPMINUS1}{
p_{-1}=-2\;,
} 
\eqn{GOTOTHERCONSTRAINT}{
2\left(p_{0}q_{0}+q_{1}\right)+q_{1}\left(p_{0}^2+p_{1}\right)=0\;.
}
These relations are indeed satisfied for all values of $d$. As a result, no logarithmic terms appear, and both
solutions are regular at $z = z_f$.

 It should also be noted that the coefficients $c_{1},c_{2}$ diverge as $z_{f}$ approaches 1.  
The case $z_{f}=1$ has therefore to be examined separately. When \eno{HORPOLE} is satisfied \eno{EOMZ} becomes
\eqn{EOMFUNNY}{
Z''[z]+\frac{d-1+z^d(1-2d)}{z(z^d-1)}Z'(z)+ \left(\frac{d^2 z^{2d-2}}{(z^d-1)^2} +q^2\frac{2-d+2z^d(d-1)}{(d-2)(z^d-1)^2}\right) Z(z)
= 0\;.
}  
The solution close to the pole is $Z(z)\sim (1-z)^s$ with 
\eqn{GORDOUBLES}{
s= 1\pm \frac{i q}{\sqrt{d(d-2)}}\;.
}  
It is then convenient to define a function $\Phi$ that has a regular Taylor expansion near $z=1$, via
\eqn{ZOUBLE}{
Z(z) = (1- z)^{s} \Phi(z).
} 
Next, one observes that for 
\eqn{GOTFUNNYQD}{
q^2= -\frac{d(d-2)}{4}
}  
the coefficients in the equation for $\Phi$  are all regular at $z=1$. Therefore, 
both $iq =\pm \frac{\sqrt{d(d-2)}}{2} $ are eigenvalues.  However, since $v_{f}$ is supersonic, the perturbation must
decrease toward negative $x$ and so only $iq=+ \frac{\sqrt{d(d-2)}}{2}$ is accepted as physical.      
  
\subsubsection{Boundary conditions}

At $z\to 0$, the spacetime becomes asymptotically AdS with the boundary at $z=0$.  Close to the boundary the two independent solutions behave
as
\eqn{BOUNDARYBEH}{
Z_{1}(z) \sim \mbox{constant}  \;,\quad Z_{2}(z)\sim  z^d\;. 
}
The first solution will change the metric on the boundary.  Since we do not wish that, we impose the boundary condition
\eqn{BOUNDCOND}{
Z(0)=0\;.
} 
Following the holographic renormalization procedure \cite{deHaro:2000xn} we also find that the boundary stress energy tensor has components 
\eqn{GOTSTRESSPERT}{
T_{\m\n} = \frac{1}{16 \pi G_{N}^{[d+1]}}\lim_{z\rightarrow 0} \frac{1}{z^d} H_{\m\n}  \;. 
}
Close to the horizon the two independent solutions are the ingoing and outgoing waves:
\eqn{HORIZONBEH}{
Z_{in}(z) \sim (1-z)^{\frac{1}{d}\frac{iqv}{\sqrt{1-v^2}}} \;,\quad Z_{out}(z) \sim (1-z)^{-\frac{1}{d}\frac{iqv}{\sqrt{1-v^2}} }\;.
}
The terminology comes about because of the way the corresponding (time-dependent) perturbations behave in the unboosted black brane 
background. Note that we need these expressions at imaginary values of $q$ (as appropriate for perturbations decaying in space), so an
analytical continuation is implied. We impose the infalling boundary condition at the horizon (that is choose the first of 
the asymptotics \eno{HORIZONBEH}). For the specifics of whether the chosen boundary conditions set a well posed problem, an interested 
reader can consult section (4.2) of \cite{Khlebnikov:2010yt}.  

\begin{figure}
\centering
\subfigure[The Taylor series around the boundary $z=0$ and the horizon $z=1$ are matched at an intermediate point.  The location of the spurious pole $z=z_{f}$ is not taken into account. When $z_{f}>1$ there is no problem as the spurious pole does not lie in the physical region.  When $z_{f}<1$ the spurious pole is in the physical region.  Interestingly it does not affect the result of the numerical matching as long as it is not very close to the horizon.  When the two poles coincide a new Taylor series has to be developed.]{
\includegraphics[width=6cm]{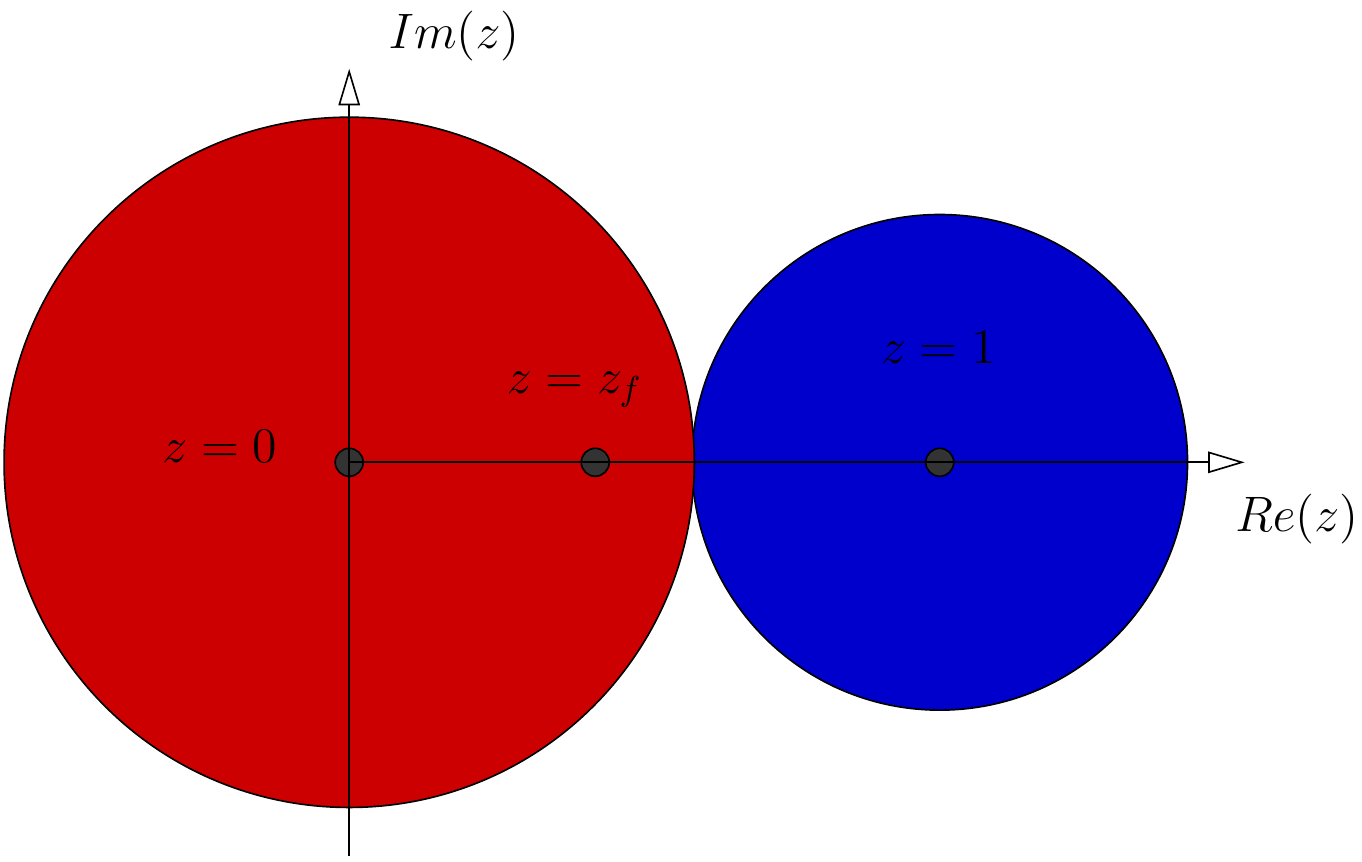}
\label{2Poles}
}
\subfigure[When the spurious pole is in the physical region we can develop three different series, around $z=0,z=z_{f},z=1$.  The series around $z=z_{f}$ has two free parameters since there is no boundary condition to take into account. It is matched with the other two series at
intermediate points $z_1$ ($0<z_1<z_{f}$) and $z_{2}$ ($z_{f}<z_2<1$). The result for $q$ is numerically indistinguishable from the method in \eno{2Poles}.]{
\includegraphics[width=6cm]{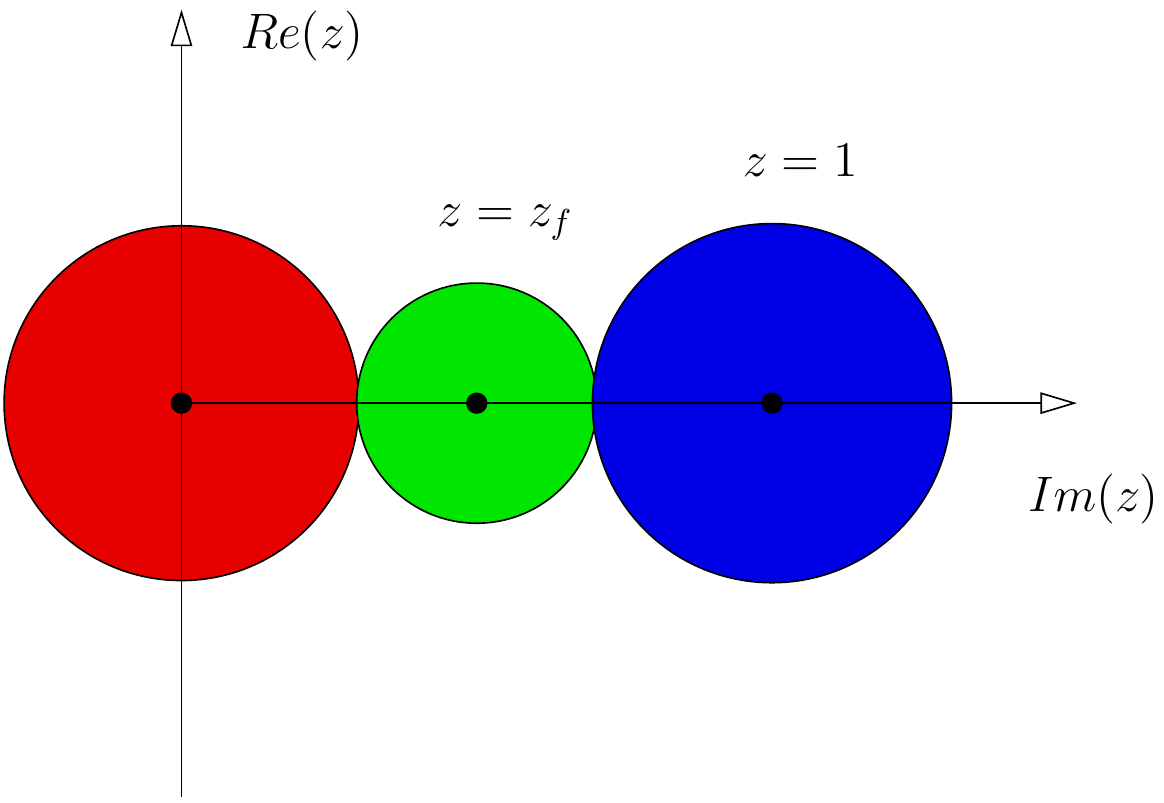}
\label{3Poles}
}
\subfigure[A third method used for determining $q$ involves expanding around $z=1$ to a low order in Taylor series and then 
integrating numerically from the horizon.  The spurious pole has to be displaced in the complex plane for the numerical integration 
to be reliable.  The sign of $\e$ has no effect and, for sufficiently small $|\e|$, the results agree well with those 
of the previous methods.]{
\includegraphics[width=6cm]{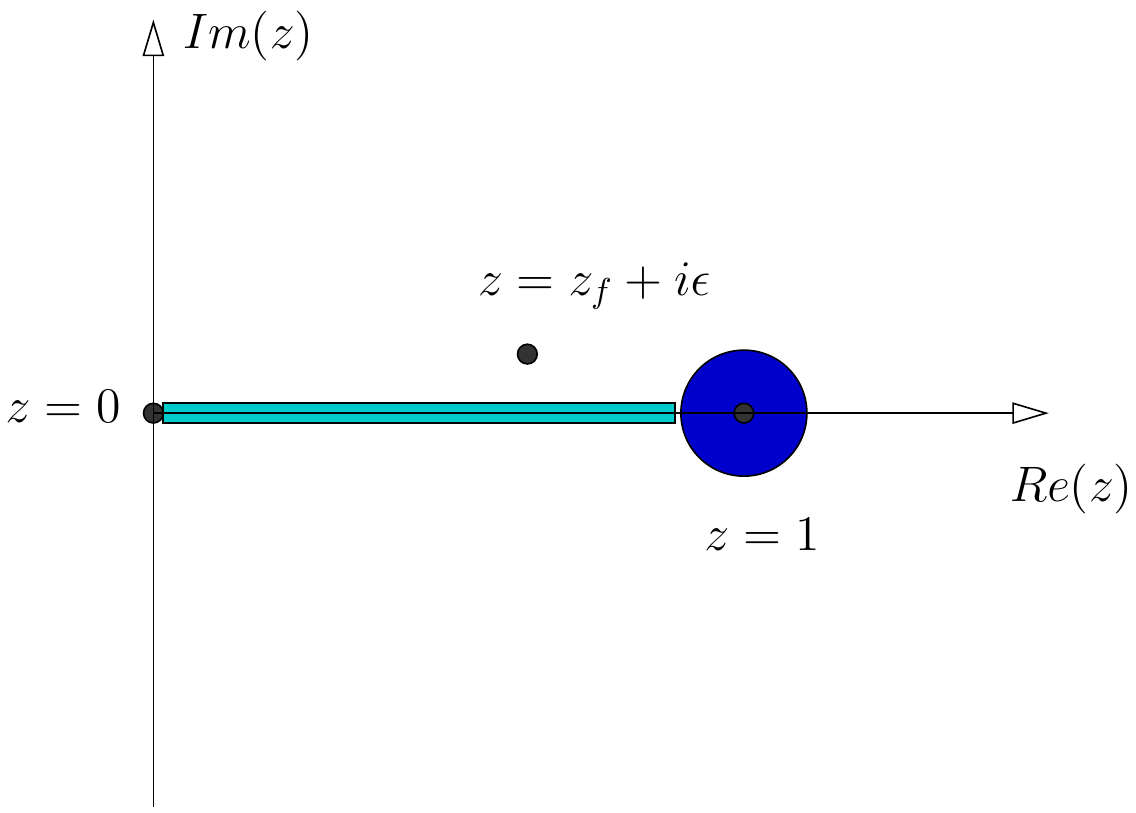}
\label{NumIntegration}
}
\label{NicePoles}
\caption[]{Different methods for determining $q$}
\end{figure}

We set up the numerical problem by looking for the solution as a Taylor expansion near the boundary and, after 
peeling off the singular part as in \eno{ZOUBLE}, a Taylor expansion near the horizon. Matching the logarithmic derivatives
of these two expansions at an intermediate point determines the eigenvalue $q$. The coefficients of the expansion
near the horizon grow when the spurious pole $z_f$ approaches 1 (cf. \eno{GOTC1} and \eno{GOTC2})), which requires us to keep a
large number of terms in the expansion (we go up to around thirty). This procedure can be further checked in two ways. 
First, since we have seen that the solution is regular at $z_f$, we can develop a third expansion near $z = z_f$.
We then match this expansion with the expansions near $z=0$ and $z=1$ at two intermediate points and find that the result 
coincides to five significant digits with result from the previous procedure.  Second, we can use the shooting method 
(integrate numerically \eno{EOMZ} from the horizon to the boundary). For this, we circumvent the spurious pole by displacing
it into the complex plane ($z_f \to z_{f} +i \e$). For a sufficiently small $\epsilon$, the results agree well with those
of the other two methods. The three methods we have used are summarized graphically
in figures \ref{2Poles},\ref{3Poles},\ref{NumIntegration}.   

\subsection{Quasinormal modes for special values of the boost velocity}


\subsubsection{Speed of sound}

When $v=\sqrt{\frac{1}{d-1}}$, \eno{EOMZ} simplifies and $q=0$ is an eigenvalue.  The two independent solutions are 
\eqn{GOTSOUNDWAVE}{
Z(z) = z^d \;, \quad Z(z) = 4+z^d \log(1-z^d)\;.
} 
Only the first one satisfies the boundary conditions.  

\subsubsection{$v=\sqrt{\frac{d}{2(d-1)}}$}

For $v=v_{f}=\sqrt{\frac{d}{2(d-1)}}$ the locations of the horizon and of the spurious pole coincide.  
The eigenvalue for this case has been found in sec.~\ref{sec:close_to_pole}:
\eqn{GOTFUNNYQ}{
iq = \frac{\sqrt{d(d-2)}}{2}\;.
} 
It is interesting to note that, for this value of $q$, the boundary exponent \eno{GORDOUBLES}, valid for the case
$z_f = 1$, coincides with the exponent in \eno{HORIZONBEH}, derived originally for $z_f \neq 1$. 
One consequence of this is that
the curve $q(v)$ is continuous at $v=v_f$.\footnote{We are talking here about the ``main'' branch of $q(v)$, which
corresponds to the slowest-decaying component of a perturbation. This is the branch we are primarily interested in,
although occasionally we will also present results for faster-decaying branches.}

For $d=3$, the wavefunction $Z(z)$ corresponding to this value of $q$ can be written in terms of a hypergeometric
function:
\eqn{GOTZd3}{
Z(z)= C_{1}Z_{3d1} +C_{2}Z_{3d2}\;,}
where
\eqn{GOTZ3d1}{
Z_{3d1}(z)= (z+2)\left(z^2+z+1\right)^{1/4}\sqrt{z^3-1}\left(\frac{2i \sqrt{3}z +i\sqrt{3}+3}{3-i\sqrt{3}-2i\sqrt{3}z}\right)^{i\sqrt{3}/4}\;,
}
\eqn{GOT3d2}{
Z_{3d2}(z)=&-\frac{1}{6}(-1)^{5/6}e^{\frac{1}{2}(\sqrt{3}-2i)\arctan(\frac{2z+1}{\sqrt{3}})}\left(z^2+z+1\right)^{1/4} \sqrt{z^3-1} \cr 
&\Bigg( -\frac{4(\sqrt{3}+i)\sqrt{(z^2+z+1)(2i\sqrt{3}z+i\sqrt{3}+3)}}{\sqrt{3-i\sqrt{3}-2i\sqrt{3}z}} \cr 
&+\frac{3(z+2)\left(1+i\sqrt{3}-2z(1-i\sqrt{3}+z)\right)}{z^2+z+1} \;_{2} F_{1} (1,\frac{1-i\sqrt{3}}{2},\frac{3-i\sqrt{3}}{2};1-\frac{6i}{3i+\sqrt{3}+2\sqrt{3}z})\Bigg) \;,
}
\eqn{C2C23d}{
C_{2} =C_{1} \frac{3 e^{-\frac{\pi}{2\sqrt{3}}} }{2-3 \;_{2}F_{1}(1,\frac{1-i\sqrt{3}}{2},\frac{3-i\sqrt{3}}{2},-\frac{i(\sqrt{3}-i)}{2})}\;.
}
It can be directly checked that this wavefunction satisfies both boundary conditions.
  
In four dimensions $Z(z)$ was determined in \cite{Khlebnikov:2010yt} to be 
\eqn{GOTZd4}{
Z(z) =-Re(y(z)) +\tilde{C}_{1}Im(y(z))\;,
}
\eqn{GOTyd4}{
y(z)= z^2 \sqrt{1-z^4} \left(\frac{1+z^2}{z^2}\right)^{\frac{1+i}{2}} \;_{2}F_{1} \left(\frac{3+i}{2},\frac{-1+i}{2};1+i;1+\frac{1}{z^2} \right)\;,}
where the constant $\tilde{C}_{1}$ is numerically determined to be $\tilde{C}_1 =0.38898$.  
In six dimensions we were not able to find a simple wavefunction.

\subsubsection{The ultrarelativistic regime $\g \rightarrow \infty$}\label{ULTRA}

When the boost velocity approaches the speed of light the location of the spurious pole approaches the boundary. 
An {\em estimate} for $q(v)$ in this case can obtained by using a small number of terms in the \eno{ZFUNSERIES}
and imposing the boundary condition $Z(0)=0$. Retaining only the first three terms (corresponding to the 
coefficients $c_{0},c_{1},c_{2}$), we obtain the estimate
\eqn{GOTRELQ}{
q^2_{est} =- \frac{(d-2)(z_{f}^d-1)(z_{f}^d(4+d(d-3))-4)}{2z_{f}^2 d}\;.
}
Substituting the value of $z_f$ from \eno{GOTFUNNYPOINT} and retaining only the leading terms in the limit
$v\to 1$, we obtain 
\eqn{GOTRELQV}{
iq_{est} \sim - \g^{2/d} \left(\frac{d-2}{2(d-1)}\right)^{1/d} \left(\frac{2(d-2)}{d}\right)^{1/2}  \;.
}
The higher-order terms in the expansion \eno{ZFUNSERIES} are non-negligible and significantly change the
numerical coefficient in front of $\g^{2/d}$. In what follows, 
we determine that coefficient numerically. However, the $\g^{2/d}$ scaling given by this simple argument 
is correct.

A more rigorous way to obtain this scaling is the following.  Numerical solution of \eno{EOMZ} suggests that when $v\rightarrow 1$ the eigenvalues $q$ become large and the maxima of the wavefunctions are close 
to  $z=z_{f}\rightarrow 0$.  One then is compelled to ignore factors of $z$ compared to unity in the equation of motion 
for $Z(z)$ \eno{EOMZ}.  This changes the functions $P(z),Q(z)$ to 
\eqn{GOTPRESCALED}{
P(z)=\frac{2(d-1)^2-\g^2 z^d (3d^2-7d+2) }{z(z^d \g^2 (d-2)-2(d-1))}\;,
}     
\eqn{GOTQRESCALED}{
Q(z)=-\frac{d^2 (d-2) \g^2 z^{2d-2}}{(z^d \g^2 (d-2)-2(d-1))}+q^2 (\g^2 z^d -1))\;.
}
We can now define a new variable 
\eqn{XDEF}{
x= \frac{z}{z_{f}}=\g^{2/d} \frac{z(d-2)^{1/d}}{\left(2(d-1)\right)^{1/d}} \;,
}
and take the limit $\g \rightarrow \infty$ keeping $x$ fixed.  Equation \eno{EOMZ} transforms into 
\eqn{GOTZRESCALED}{
Z''(x) - Z'(x) \frac{(3d -1)x^d+1-d }{x(x^d-1)} + Z(x) p^2 \left(-1+2\frac{d-1}{d-2}x^d \right)=0\;.}
where
\eqn{GOTPQRESCALED}{
p^2= z_{f}^2 q^2 \;.
}
Nothing in \eno{GOTZRESCALED} depends on $\g$.  The eigenvalue $p$ for this problem is therefore a simple number.  The scaling for $q$ then immediately follows from \eno{GOTPQRESCALED}: 
\eqn{GOTQSCALING}{
q \sim \frac{1}{z_{f}} \sim  \g^{\frac{2}{d}}\;, \quad (\g \rightarrow \infty)\;.
}
 The physical region $0<z<1$ gets mapped to $0<x<\infty$ with Dirichlet conditions at both ends.  The spurious pole at $x=1$ can be circumvented in numerical integration by moving it in the complex plane to $x=1+i \e$ and then taking the limit $\e\rightarrow 0$.  Numerically, we find the following for physically interesting dimensions:
\eqn{GOTQRd3}{
iq = 1.019 \g^{2/3} \;, \quad d=3\;,
}
\eqn{GOTQRd4}{
iq = 1.895 \g^{1/2} \;, \quad d=4\;,
}
\eqn{GOTQRd6}{
iq =  3.62\g^{1/3} \;, \quad d=6\;.
}

\subsection{Quasinormal modes for the boosted black hole in $AdS_{4}$}\label{ADS4}

One physically interesting case is $d=3$.  The relevant SCFT is the Chern-Simons theory described by \cite{Aharony:2008ug} and 
referred in the literature as ABJM, heated to a temperature $T$.  The transport coefficients up to the 
second order have been calculated in \cite{VanRaamsdonk:2008fp}.  Here, we find the values of $q(v)$ for various 
$v$ in the region $0<v<1$. The highest $\g$ factor for which we determine $q$ is $\g=35$.  
Some interesting properties of the $q(v)$ curve are as follows.

For $v=0$ we find $iq=-2.1486$.  This is the mass of the lowest-mass glueball in $QCD_{2}$.\footnote{We define $QCD_{2}$ as a theory obtained by a Scherk Schwarz compactification \cite{Scherk:1979zr},\cite{Scherk:1978ta} of ABJM down to $1+1$ dimensions.}  It should be pointed out, though,
that there are no shock wave solutions for which $v=0$ (on either side), so this quasinormal mode is not directly relevant 
to our problem.

For $v=1/\sqrt{2}$ (the speed of sound) we find  $iq=0$, as expected. For small departures from the speed of sound,
$v\sim 1/\sqrt{2} +\d v$, we find, to the linear order in $\delta v$,
\eqn{GOTHYDROQ3NUM}{
iq(v) = 4.24264 \; \d v = 3\sqrt{2}\; \d v\;.
}
This is in agreement with the hydrodynamic result of section \eno{FIRSTORDER} (provided we rescale \eno{GOTHYDROQ3NUM} 
by a factor $z_{h}=\frac{4\pi T}{3}$ ).  
For $v=v_{f}=\sqrt{\frac{3}{4}}$ we find numerically $iq =0.86602=\sqrt{3}/2 $ in perfect agreement with our analytical
result \eno{GOTFUNNYQ}.  
We also numerically explore the relativistic limit and find agreement with the analytical computations of Section ~\ref{ULTRA}. 
The numerical curve $q(v)$ is shown in figure \eno{qh3dFig}.

It should be noted here that a simple approximation reproduces the curve quantitatively up to a $5\%$ accuracy, 
with the largest error coming from the hydrodynamic region where $iq$ is small.  The approximation is given by 
 \eqn{GOTQ3APPROX}{
 iq = \g^{2/3} \frac{Av+B}{Fv+G}\;,
 }    
 where 
 \eqn{GOT3ABFG}{
 A=1\;, \quad B= -1/\sqrt{2}\;, \quad F= 0.1539\;, \quad G=1/(2.1486 \sqrt{2})\;.
 }
The values of $A,B,F,G$ are determined by requiring that the curve matches the numerics at the three points 
$v=0,\frac{1}{\sqrt{2}},\sqrt{\frac{3}{4}}$.  The function \eno{GOTQ3APPROX} does not change under the rescaling of all the coefficients so only three constants are independent.  
If one boosts back to the frame where the shock front is moving, this curve can give a good approximation for the location of the quasinormal modes on the complex $\oo$ plane when $\oo' = - v q' $. Such a determination might be relevant if an effective theory of hydrodynamics of the type described in \cite{Lublinsky:2009kv} needs to be pursued.   Other values of $A,B,F,G$ can be obtained if the point $v=0$ is replaced by the asymptotic behavior at $v\rightarrow 1$.  
This produces more accurate results for the region $v>v_{f} = \sqrt{\frac{3}{4}}$.

\begin{figure}
\centering
\subfigure[Values of $Im(q)$ versus the velocity.  The blue curve is the main branch, the green the first upper branch, and the dashed red the hydrodynamical approximation. In the hydrodynamical approximation $Im(q)$ diverges at $v=0$.  
It is only a good approximation close to $v=v_{s}= \sqrt{\frac{1}{2}}$]{
\includegraphics[width=10cm]{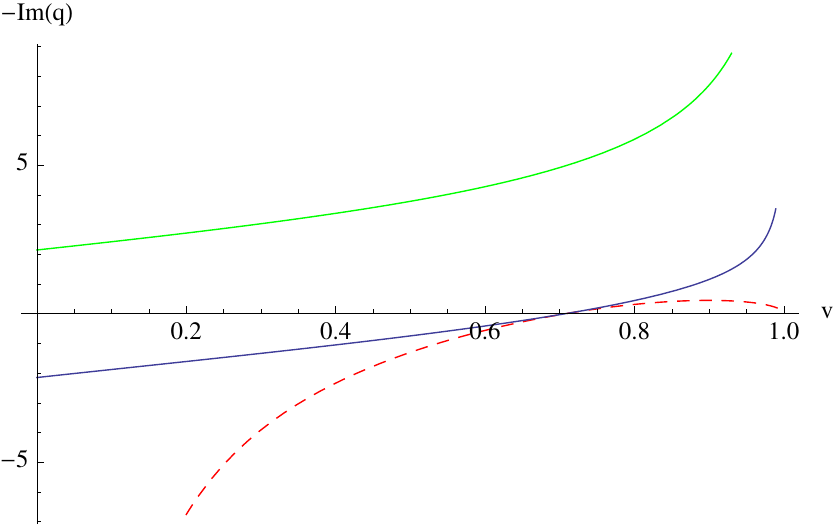}
\label{qh3dFigv}
} 
\subfigure[Values of $Im(q)$ versus $\g=\frac{1}{\sqrt{1-v^2}}$.  Both the main (blue) and the upper (green) branch diverge 
at large $\gamma$ as $\g^{2/3}$. We have determined $Im(q)$ beyond the values shown here up to $\g=35$.  
The hydrodynamical approximation (red) gives vanishing $q$ for large $\g$.]{
\includegraphics[width=10cm]{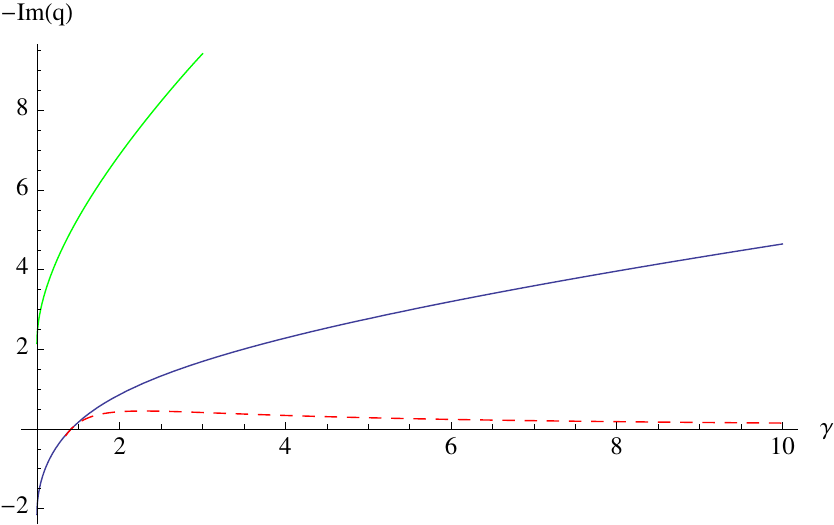}
\label{qh3dFigg}
}
\caption[]{Numerical values of $q$ for different velocities in $d=3$.}
\label{qh3dFig}
\end{figure}

\subsection{Quasinormal modes for the boosted black hole in $AdS_{7}$}\label{ADS7}

 Another physically interesting case is $d=6$.  This theory is dual to a stack of $N$ $M5$ branes.  The transport coefficients for this theory were obtained in \cite{Haack:2008cp,Bhattacharyya:2008mz} as a part of the calculation for generic dimension.   The numerical method we have used to determine $q$ is the expansion in Taylor series around the boundary and the horizon and matching of the logarithmic derivatives at a point between $0$ and $1$.  The numerical result up to $\g=10$ is shown in figure \eno{qh6dFig}..  

For $v=0$, we find that the lowest glueball mass for the dimensionally reduced theory is given by $iq=-2.703$.  For the speed of sound ($v=1/\sqrt{5}$)
$iq=0$, and for small deviations $v= 1/\sqrt{5}+\d v$ we find 
\eqn{GOTHYDROQ6NUM}{
iq = 6.7082 \;\d v = 3\sqrt{5}\; \d v\;,
}        
in agreement with \eno{GOTQLIN} (remembering that $z_{h}=\frac{2\pi T}{3}$ in this case).  For the case of the spurious pole coinciding with the horizon the numerical result agrees with the analytically predicted one. Finally the relativistic regime is found to be well described by \eno{GOTFUNNYQ}.

There is a simple approximation for the $q(v)$ curve similar to \eno{GOTQ3APPROX}:
 \eqn{GOTQ6APPROX}{
 iq = \g^{1/3} \frac{Av+B}{Fv+G}\;,
 }    
 where now
 \eqn{GOT6ABFG}{
 A=1\;,\quad B= -1/\sqrt{5} \;, \quad F= -0.0125 \;, \quad G= 1/(2.7034\sqrt{5})\;.
 }
 
\begin{figure}\label{qh6dFig}
\centering
\subfigure[Values of $Im(q)$ versus the velocity for $d=6$.  The blue curve is the main branch and the dashed red the hydrodynamical approximation.  In the hydrodynamical approximation $Im(q)$ diverges at $v=0$.  It is only a good approximation close to $v=v_{s}= \sqrt{\frac{1}{5}}$.  We do not show here the upper branch.]{
\includegraphics[width=10cm]{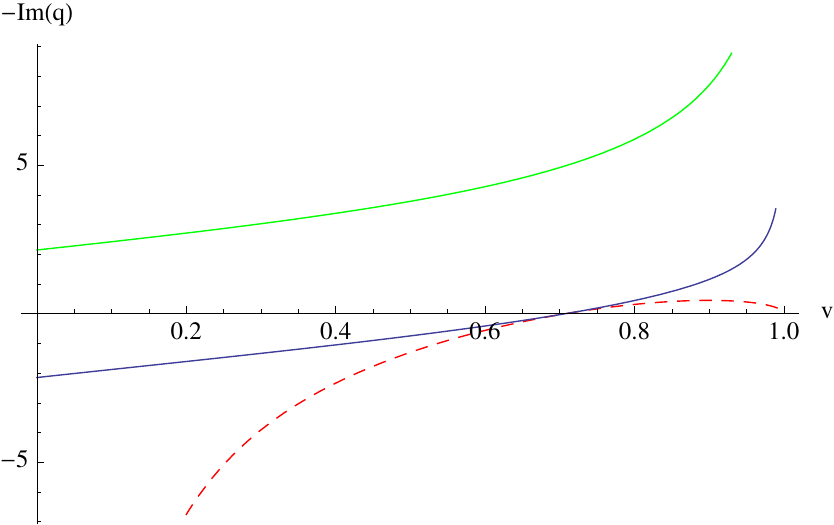}
\label{qh6dFigv}
}
\subfigure[Values of $Im(q)$ versus $\g=\frac{1}{\sqrt{1-v^2}}$.  The main (blue) branch diverges as $\g^{1/3}$.  The hydrodynamical approximation (red) gives vanishing $q$ for large $\g$]{
\includegraphics[width=10cm]{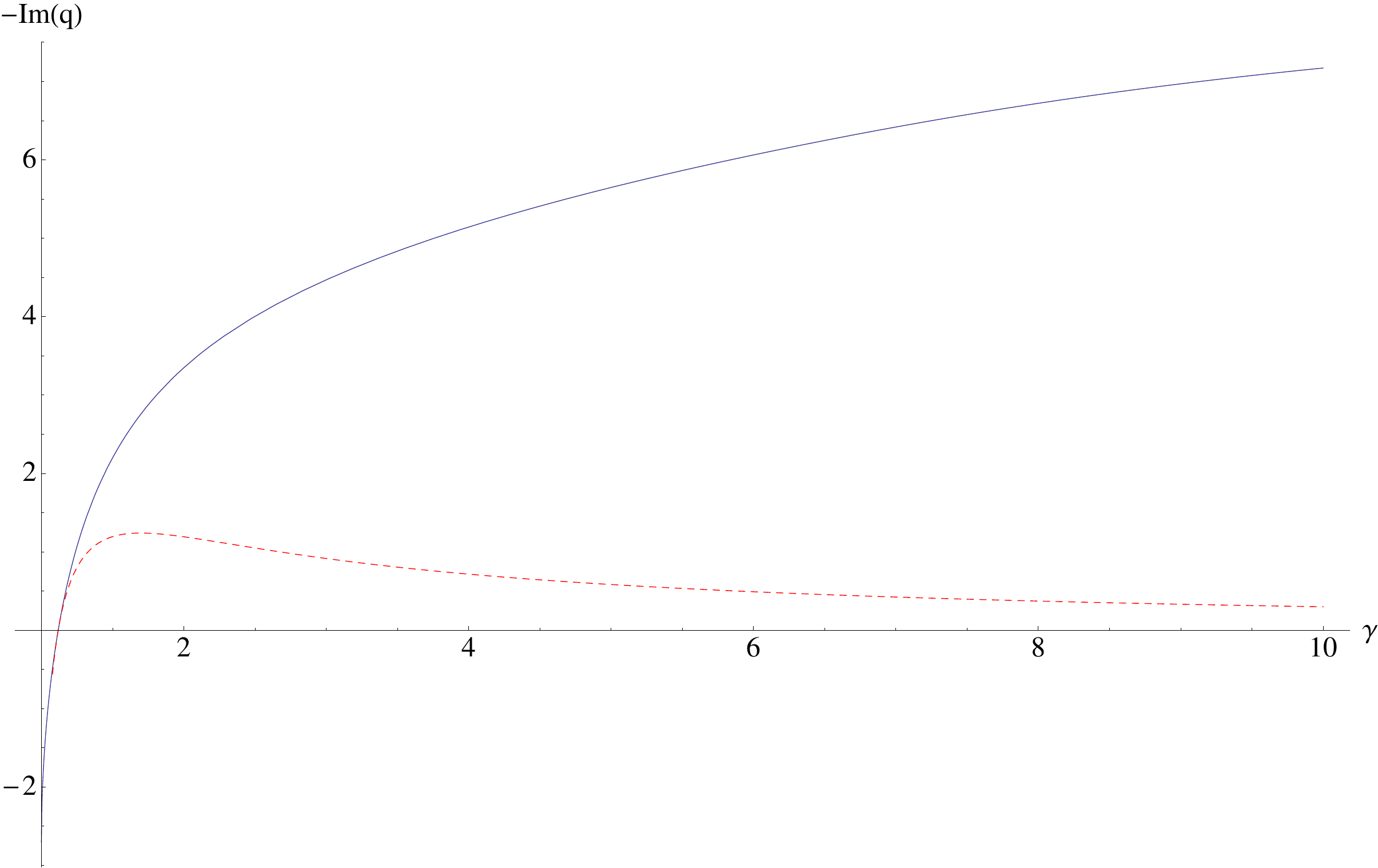}
\label{qh6dgFigg}
}
\caption[]{Numerical values of $q$ for different velocities in $d=6$}
\end{figure}

 
\section{Properties of the second-order metric}\label{KILLING}
\subsection{Expression for the metric}

We now turn to the case of weak shocks, where we would like to use  gradient expansion  to infer properties of black holes in the dual
gravity theory. In this section, as well as the next one, 
we restrict ourselves to the case $d=4$.

For any solution of fluid mechanics that has small gradients of the fluid velocity 
and temperature, the dual bulk metric
can be reconstructed order by order in the gradient expansion \cite{Bhattacharyya:2008jc}. In our previous paper \cite{Khlebnikov:2010yt}, 
we have found the first few terms in this expansion for planar weak shocks in 
the ${\cal N}=4$ fluid that extends indefinitely in the direction of propagation
(the $x$ direction). The expansion parameter is in this case $\uinf$, the first-order 
variation of the asymptotic 4-velocities relative to the speed of sound (cf.
sec.~\eno{FIRSTORDER}). This parameter controls the
size of the gradient terms.

Here, we discuss some properties of the metric computed to the second order in $\uinf$.
It has the following structure:
\be
g_{MN}(r,x) = \left( \begin{array}{ccccc} g_{tt}(r,x) & u^0(x) & g_{tx}(r,x) & 0 & 0  \\
u^0(x) & 0 & - u(x) & 0 & 0 \\
g_{tx}(r,x) & - u(x) &  g_{xx}(t,x) &  0 & 0 \\
0 & 0 & 0 & r^2 - \half s_T(r,x) & 0 \\
0 & 0 & 0 & 0 & r^2 - \half s_T(r,x) \end{array} \right) .
\label{met}
\ee
We denote by $u$ the component of 4-velocity of the fluid (relative to the shock) in the 
$x$ direction and by $u^0$ the corresponding temporal component $u^0 = \sqrt{1+u^2}$.
The function $s_T(r,x)$, which appears in the transverse components of the metric, is of 
the second order in $\uinf$ and is given by eq.~(\ref{sT}) below.

Each of the metric functions in (\ref{met}) includes a piece that has the same form 
as the equilibrium metric $g^{(0)}_{MN}$ but now with slowly varying parameters, plus
a ``standalone'' second-order correction:
\[
g_{MN}(r,x) = g^{(0)}_{MN}[r, u(x), T(x)] + H_{MN}(r,x) \; .
\]
Explicitly, the equilibrium metric is
\be
g^{(0)}_{MN}[r,u,T] = \left( \begin{array}{ccccc} -r^2 + \frac{r_T^4 (u^0)^2}{r^2} &
u^0 & - \frac{r_T^4 u^0 u}{r^2} & 0 & 0 \\
u^0 & 0 & - u & 0 & 0 \\
- \frac{r_T^4 u^0 u}{r^2} & - u & r^2 + \frac{r_T^4 u^2}{r^2} & 0 & 0 \\
0 & 0 & 0 & r^2 & 0 \\
0 & 0 & 0 & 0 & r^2 \end{array} \right) ,
\label{met0}
\ee
with $r_T \equiv \pi T$. The slowly-varying parameters (to the second order) are
\ba
u(x) & = & \frac{1}{\sqrt{2}} + u_{(1)}(x) + u_{(2)}(x) \; , \label{ux} \\
T(x) & = & T_{(0)} \left( 1 - \frac{\sqrt{2}}{3} u_{(1)}(x) \right) + T_{(2)}(x) \; ,
\label{Tx}
\ea
where 
\be
u_{(1)}(x) = - u_{\infty} \tanh \frac{4\pi T_{(0)} \uinf x}{3} \; ,
\ee
while $u_{(2)}$ and $T_{(2)}$ are related as follows:
\be
\pi T_{(2)} = \frac{1}{3} u_{(1)}' - \frac{\sqrt{2}}{3} \pi T_{(0)} u_{(2)} \; .
\label{T2}
\ee
Explicit expressions for $u_{(2)}$ and $T_{(2)}$ individually are also available
\cite{Khlebnikov:2010yt}, but will not be needed in what follows.

Note that $u_{(1)}$ depends on two constants,  
$T_{(0)}$ and $\uinf$, the first of which can be thought as the ``base'' temperature
of the fluid, and the second as the strength of the shock. The gradient expansion, 
in the present case, is an expansion in powers of $\uinf \ll 1$. The value $1/\sqrt{2}$ for 
the leading term in (\ref{ux}) reflects the fact that weak shocks are 
only possible for velocities close to the speed of sound.

Finally, the ``standalone'' corrections are
\begin{eqnarray}
H_{tt}(r,x) & = & \half s_T(r,x) \; , 
\label{Htt} \\
H_{tx}(r,x) & = & \frac{1}{\sqrt{3}} r u_{(1)}'(x) - \frac{\sqrt{3}}{2} s_T(r,x) \; , \\
\label{Htx}
H_{xx}(r,x) & = & -\frac{2}{3} r u_{(1)}'(x) + \frac{3}{2}  s_T(r,x) \; .
\label{Hxx}
\ea
where
\be
s_T(r,x) = \frac{r^2}{3\pi T_{(0)}} F_1 \left (\frac{r}{\pi T_{(0)}} \right) u_{(1)}'(x)  
\label{sT}
\ee
with
\be
F_1(y) = \ln \frac{(1+y^2)(1+y)^2}{y^4} - 2 \arctan y + \pi \, ,
\ee
and primes denote derivatives with respect to $x$.

\subsection{Location of the horizon}
Define the horizon of the second-order metric (\ref{met}) as the surface
\be
r = r_H(x) \, ,
\label{hor}
\ee
at which the normal vector
\be
V_M = (0, 1, -r_H', 0 , 0) 
\label{normal}
\ee
becomes null. This is an apparent horizon defined similarly to the ones in 
\cite{Bhattacharyya:2008xc,Booth:2011qy}.  We expect that, just as the metric itself,
$r_H$ is a sum of the slowly varying
\be
r_T(x) \equiv \pi T(x)
\ee
and a standalone second-order term. The standalone term is sourced by $r_T'$ and so
is $O(\uinf^2)$. In addition, it varies slowly with $x$, so its derivative is 
$O(\uinf^3)$. Thus, to the second order order inclusive 
\be
r_H'(x) = r_T'(x) = - \frac{\sqrt{2}}{3}  \pi T_{(0)} u_{(1)}'(x) \; .
\label{rHprime}
\ee
The null condition becomes
\be
g^{rr}(x) - 2 u r_T'(x) = 0 \, ,
\label{null_cond}
\ee
where to the required accuracy we can set $u = 1/ \sqrt{2}$.

A bit of algebra yields
\be
g^{rr}(r,x) = r^2 \left( 1 - \frac{r_T^4(x)}{r^4} \right) - \frac{2}{3} r u_{(1)}'(x) \, ,
\label{grr}
\ee
for $r\to r_H(x)$ (and to the second order in $\uinf$).
When this is used in (\ref{null_cond}), there is 
a curious cancellation between the last terms in (\ref{null_cond}) and (\ref{grr}).
As a result, the equality $r_H \approx r_T$ is good to the second order inclusive, i.e.,
\be
r_H(x) = \pi T(x) + O(\uinf^3) \; .
\label{rHx}
\ee
In other words, at the second order, $r_H(x)$ tracks point by point the slowly varying 
temperature (\ref{Tx}).

\subsection{Killing vectors}
The second-order metric (\ref{met}) has Killing vectors corresponding to translations
in $t$, $y$, and $z$. The zeroth-order metric, given by (\ref{met0}) with
$x$-independent parameters, has in addition
a Killing vector corresponding to translations in $x$. For that zero-order metric,
a linear combination of $\partial_x$ and $\partial_t$ with appropriate coefficients yields
the Killing vector 
\be
\chi^M_{(0)}[u] = (u^0, 0, u, 0, 0) \; ,
\label{chi0}
\ee
for which the horizon of (\ref{met0}) at $r=r_T$ is a Killing horizon, i.e.,
$\chi^M_{(0)}$ is normal to it at every point.

In general, existence of a Killing vector normal to the horizon has important consequences 
for the physics of a black hole. 
In particular, it can be used (with suitable additional assumptions) to prove 
that the surface gravity of the horizon is a constant \cite{Wald:1984rg}. 
Thermodynamics of black holes identifies
the surface gravity with temperature, so existence of a Killing vector normal to 
the horizon can be thought of as the condition that the black hole is in equilibrium. 
Moreover, for stationary black holes with {\em compact} horizons, existence of such 
vectors can often be established using Hawking's theorem \cite{Hawking:1971vc,Hawking:1973uf}.
This reinforces the intuition 
that stationary metrics with compact event horizons correspond to equilibrium black 
holes.

The second-order  metric (\ref{met}) is stationary but the horizon (\ref{hor}) is 
not compact.
On physical grounds, we do not expect our black hole (or, rather, black brane)
to be in equilibrium; indeed, it is dual to an expressly nonequilibrium, 
entropy-producing process---propagation of a shock wave. 
This means that the condition that the horizon is compact, in the statement of 
Hawking's theorem,
is not a mere technicality. Indeed, recall that
the time-independence of the metric (\ref{met}) comes about
because we can boost to the frame where the shock wave is at rest. This will not
be possible, for instance, in global coordinates, where the boundary of AdS is
a sphere. In that case, the horizon is compact but no longer time-independent: we expect 
it to contain a solitary spherical wave that propagates, say, from north to south and 
eventually equilibrates near the south pole.

Even though the metric (\ref{met}) does not satisfy the compactness condition, and so
Hawking's theorem cannot be used to establish existence of Killing vector normal to
the horizon, it is interesting to ask if such a vector
nevertheless exists. That would mean that the metric has a hidden symmetry 
that we have up to now failed to detect. We therefore ask if there is a deformation of 
(\ref{chi0}),
\be
\chi^M(r,x) = \chi^M_{(0)}[u(x)] + k^M(r,x)
\label{kM}
\ee
that satisfies the Killing equation
\be
\nabla^M \chi^N + \nabla^N \chi^M = 0
\label{keq}
\ee
up to and including $O(\uinf^2)$ terms. Note that (\ref{kM}) has the same structure 
as our earlier expansions in $\uinf$; in particular, it starts with 
the equilibrium result now considered as a function of a slowly-varying parameter.
Since $\chi_{(0)}$ remains a Killing vector
to the first order and, up to a trivial rescaling, is the only such vector that is
null at the horizon, we may assume that $k^M$ in (\ref{kM}) is $O(\uinf^2)$. This 
allows us to neglect derivatives of $k^M$ with respect to $x$:
\be
 k^M(r,x) \approx  k^M(r) \, .
\ee
Substituting (\ref{kM}) into (\ref{keq}) and expanding to the second order in $\uinf$, 
we find that vanishing of the $(yy)$ and
$(zz)$ components of the Killing equation requires $k^r=0$, but then the $(rr)$ 
component is explicitly nonzero:
\be
\nabla^r \chi^r = \frac{2 u}{r^2} r_T^3 r_T' \neq 0 \; .
\ee
We conclude that there is no Killing vector of the form (\ref{kM}) at the second order.

\subsection{Expansion of the horizon}
The Killing equation (\ref{keq}) is equivalent to 
\be
\lie_\chi g_{MN} = 0 \, .
\label{klie}
\ee
where $\lie_\chi$ is the Lie derivative along a putative Killing vector. The absence of 
a solution at the second order means that there 
is no symmetry of the {\em entire spacetime} whose Killing vector would be normal to 
the horizon. It is then natural to ask if 
there is a symmetry of the {\em horizon alone} that may do instead. This motivates one
to consider, instead of (\ref{klie}), the quantity 
\be
h_{ab} = \lie_{\tV} \tg_{ab} \, ,
\label{hab}
\ee
where $\tg_{ab}$
is the induced metric on the horizon, and $\tV$ is a tangent vector constructed as
follows. Consider the normal (\ref{normal}). Since $V^M = g^{MN} V_N$ is tangent to 
the horizon, it can be written as the pushforward of some vector $\tV^a$ that lives in the 
tangent space (and has one fewer component than $V^M$):
\be
V^M = \frac{\partial x^M}{\partial \xi^a} \tV^a \; .
\label{pushf}
\ee
Here, $x^M$ are coordinates in the spacetime, and $\xi^a$ are coordinates on the horizon.
$\tV$ is the vector used in (\ref{hab}).
With the choice $\xi^a=(t,x,y,z)$, the relation (\ref{pushf}) becomes simple:
$\tV^a = V^a$ for all $a$.

In general, vanishing of $h_{ab}$ is perhaps the broadest sense in which a horizon can be
considered equilibrium or non-expanding.  As such, this criterion has appeared in a number 
of recent papers (for a review, see \cite{Ashtekar:2004cn})\footnote{We thank A.\; Ashtekar for bringing \cite{Ashtekar:2004cn} to our attention.}. We will now see that it is not
satisfied in our present case.  

Although, as we will see, $h_{ab}$ is nonzero already at the second order in $\uinf$, 
for use in the subsequent discussion, we compute it to the third order. For this, we
need $\tg_{ab}$ and $\tV^a$ only to the second order (all the derivatives
in the computation of $h_{ab}$ are taken in the tangent space, so each produces
an additional power of $\uinf$).
The nonzero components of the induced metric are
\ba
\tg_{tt}(x) & = & r_H^2(x) u^2(x) + H_{tt}(r_H,x) \, , \\
\tg_{tx}(x) & = & - r_H^2(x) u^0(x) u(x)+ H_{tx}(r_H,x) + u^0 r_H'(x) \, , \\
\tg_{xx}(x) & = & r_H^2(x) u_0^2(x) + H_{xx}(r_H,x) - 2 u r_H'(x) \, , \\
\tg_{yy}(x) & = & \tg_{zz}(x) = r_H^2(x) - \half s_T(r_H,x) 
\ea
Using the expressions (\ref{Htt})--(\ref{Hxx}) and working to the second order,
we simplify the first three entries into
\ba
\tg_{tt}(x) & = & r_H^2(x) u^2(x) + \half s_T(r_H,x) \, , \\
\tg_{tx}(x) & = & - r_H^2(x) u^0(x) u(x) - \frac{\sqrt{3}}{2} s_T(r_H,x) \, , \\
\tg_{tx}(x) & = & r_H^2(x) u_0^2(x) + \frac{3}{2} s_T(r_H,x) \, .
\ea
In addition, to this order,
\be
s_T(r_H,x) =   \left(\ln 2 + \frac{\pi}{6} \right) \pi T_{(0)} u_{(1)}'(x)
\ee
and
\be
\tV^a(x) = (u^0(x), u(x), 0, 0) \; .
\label{tV}
\ee
Note that there in no standalone correction for $\tV^a$: the only second-order terms 
are those contained in the slowly varying parameters $u$ and $u^0$.

Computing the Lie derivative (\ref{hab}), we obtain the following third-order result:
\be
h_{ab}(x) = q(x) \left( 
\begin{array}{cccc} 
u^2(x) & - u(x) u^0(x) & 0 & 0 \\
- u(x) u^0(x) & [u^0(x)]^2 & 0 & 0 \\
0 & 0 & -\half  & 0 \\
0 & 0 & 0 & -\half
\end{array} \right) ,
\label{third_order}
\ee
where 
\be
q(x) = \frac{4}{3} r_H^2(x) u'(x) + \frac{1}{\sqrt{2}} s'_T(r_H,x) \, .
\label{qfunction}
\ee
The leading nonvanishing contribution to $h_{ab}$ is of the second order in $\uinf$; in
(\ref{third_order}), we have such leading term plus the next-order correction.

The leading order $h_{ab}$ satisfies the curious relation
\be
h^{(2)}_{ab}(x) = 2  \pi^2 T_{(0)}^2 \sigma_{ab} \, ,
\label{rel_to_sigma}
\ee
where
\be
\sigma_{\mu\nu} = \half P_\mu^{~\alpha} P_\nu^{~\beta} 
(\partial_\alpha u_\beta + \partial_\beta u_\alpha)
- \frac{1}{3} P_{\mu\nu} \partial_\alpha u^\alpha
\ee
with
\be
P_{\mu\nu} = \eta_{\mu\nu} + u_\mu u_\nu
\ee
determines the first dissipative correction to the stress tensor 
of a conformal fluid (in the flat 4-dimensional spacetime).  Thus, (\ref{rel_to_sigma}) 
relates the expansion of the horizon, as measured by the Lie derivative (\ref{hab}), 
to entropy production in the dual fluid.

At the next order, the simple relation of $h_{ab}$ to the stress tensor of the dual fluid
breaks down. It is worth noting, though, that the definition of entropy production at 
that order is already ambiguous \cite{Bhattacharyya:2008xc}.

We conclude this section by extracting a scalar measure of the horizon's expansion.
For that, we need to define an 
inverse $\tg^{ab}$ of the induced metric $\tg_{ab}$. Since
the vector $\tV$, eq. (\ref{tV}), is a null eigenvector of $\tg_{ab}$, such an inverse 
is uniquely defined only within the subspace orthogonal to $\tV$. However, since 
$\tV$ is also a null eigenvector of the expansion 
tensor (\ref{third_order}), the scalars we define
below will be insensitive to this ambiguity.

Consider then the projector on the subspace orthogonal to $\tV$. 
To the second order inclusive, it has the form
\be
\tP^a_{~b}(x) = \frac{1}{f[u(x)]} \left( 
\begin{array}{cccc} 
u^2(x) & - u^0(x) u(x)  & 0 & 0  \\
- u^0(x) u(x)  & [u^0(x)]^2  & 0 & 0 \\
0 & 0 & f[u(x)] & 0 \\
0 & 0 & 0 & f[u(x)]
\end{array} \right) ,
\ee
where $f[u] = (u^0)^2 + u^2$. Define the inverse metric by the relation
\be
\tg^{ab}(x) \tg_{bc}(x) = \tP^a_{~c}(x) \, .
\ee
A convenient diagonal solution is
\be
\tg^{ab}(x) = \frac{1}{r_H^2(x) f[u(x)]} \left( 
\begin{array}{cccc} 
1 & 0 & 0 & 0  \\
0 & 1 & 0 & 0 \\
0 & 0 & f[u(x)] & 0 \\
0 & 0 & 0 & f[u(x)]
\end{array} \right) .
\ee
The requisite scalars are the eigenvalues of the matrix $h^{a}_{~c} = \tg^{ab} h_{bc}$. The
latter are
$\lambda_1(x)$, 0, and $-\half \lambda_1(x)$ (the last one with multiplicity 2),
where
\be
\lambda_1(x) = \frac{1}{r_H^2(x)} q(x) \, ,
\ee
which measures the expansion of the horizon.

\section{Corrugation (in)stability and sound waves}\label{CORRUGATION}
\subsection{Matching conditions}

\begin{figure}
 \begin{center}
\includegraphics[width=5cm]{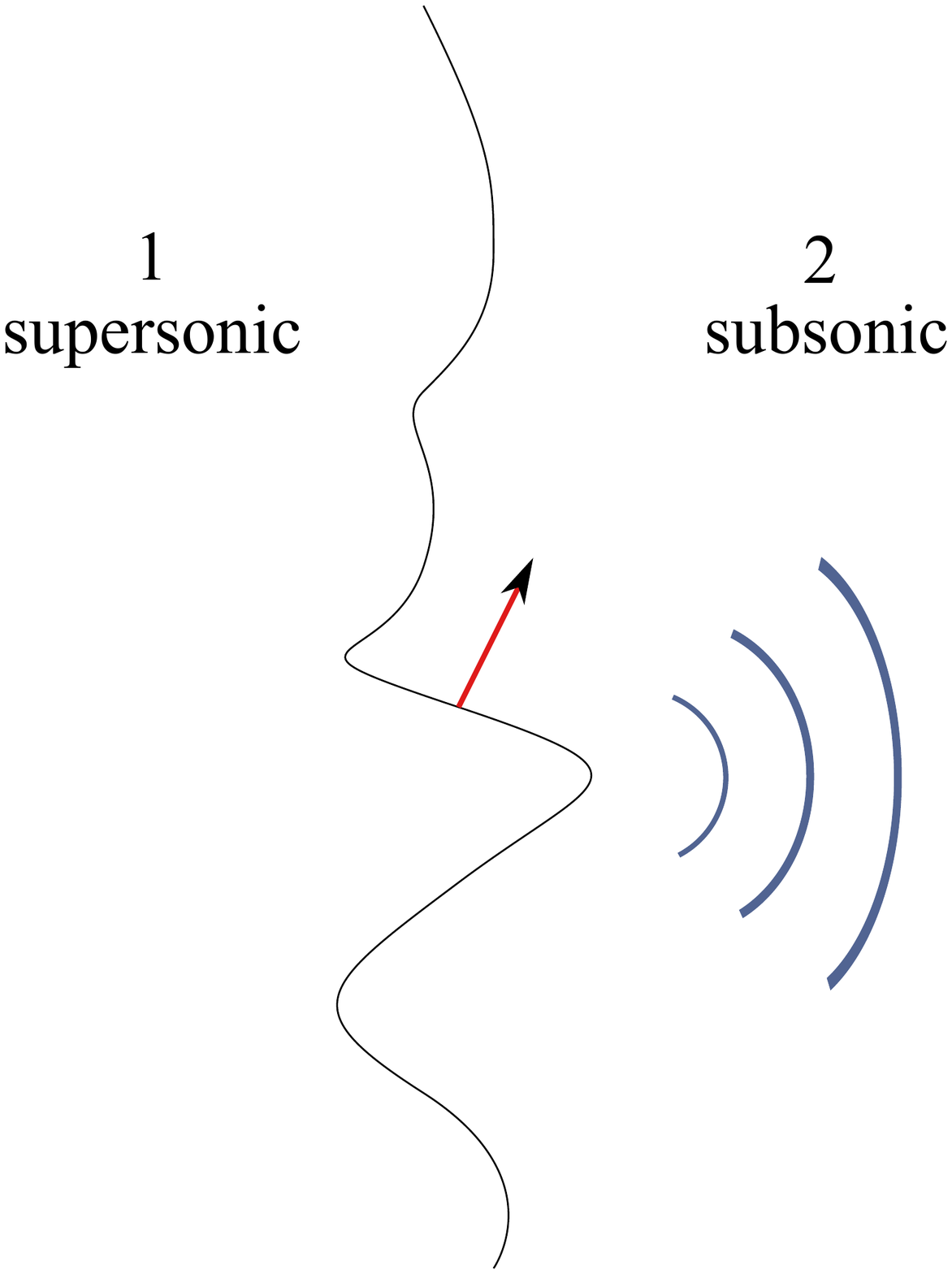}
 \end{center}
 \caption[]{A sketch of the corrugation perturbation.  In the supersonic part of the flow there is no sound or shear (entropy) wave propagating. The arrow indicates a normal to the corrugated shock front.
Conservation of the stress energy tensor across the surface is given by \eno{ACROSS}}
\label{corrugation2Fig}
 \end{figure}

 Shock waves can be unstable under oscillations of the shock interface. For small 
 perturbations, stability can be tested by a linearized approximation where the existence of exponentially growing
 modes corresponds to an unstable situation. Even if a shock is stable,
there is still the interesting possibility of a mode which is not damped.
 Once such a perturbation is established, it does not decay in time. This usually describes emission of a wave (e.g., sound) by the shock, the energy
 for such emission being provided by the shock itself.  In this section we make the linear analysis for the relativistic shock. 
 As a first step however we need to establish the matching conditions for a shock of generic shape.  If the shock is described by a surface 
 whose normal is everywhere space like and denoted by $n^\mu$ with $n^2=1$ then the matching conditions are simply
 \beq\label{ACROSS}
 n_\mu T_1^{\mu\nu} = n_\mu T_2^{\mu\nu}\;.
 \eeq
 This is just the continuity of  energy and momentum flow through the shock. For the energy momentum tensor we take the ideal fluid case
 \beq
 T^{\mu\nu} = T^4 \left(\eta^{\mu\nu} + 4 u^\mu u^\nu\right)\;,
 \eeq
 and therefore we get
 \beq 
 T_1^4\left(n^\nu+4(n\cdot u_1) u_1^\nu\right) = T_2^4\left(n^\nu+4(n\cdot u_2) u_2^\nu\right) \;.
 \eeq
 This equation can be thought as a way to determine $T_2, u_2^\nu$ from $T_1$ and $u_1^\nu$ (or vice versa). In order to do so we project the
 equation along $n^\mu$, $u_1^\nu$ and $u_2^\nu$. This gives rise to three scalar equations which can be recast as
 \beqa
 1+(n\cdot u_1)^2 + (n\cdot u_2)^2 - 8 (n\cdot u_1)^2(n\cdot u_2)^2 &=&0 \;,\\
  u_1\cdot u_2 + 2(n\cdot u_1)(n\cdot u_2)  &=& 0 \;,\\
  T_1^4\left(1+4(n\cdot u_1)^2\right)=T_1^4\left(1+4(n\cdot u_2)^2\right) \;.
 \eeqa
 The first equation gives 
 \beq
  (n\cdot u_2) = \sqrt{\frac{1+(n\cdot u_1)^2}{8(n\cdot u_1)^2-1}}
  \eeq
 and the other two can then be used to trivially obtain $(u_1\cdot u_2)$ and $T_2$. Notice that we need $(n\cdot u_1)>\frac{1}{\sqrt{8}}$. This is equivalent to saying that
 the velocity perpendicular to the interface is $v_1>\frac{1}{3}$, a constraint that we already knew.
 Now we would like to obtain the vector $u_2^\nu$.  It is clear that $u_2^\nu$ is a linear combination of $u_1^\nu$ and $n^\nu$ and it is not difficult to see that
 \beq
 u_2^\mu = \frac{3(n\cdot u_1) u_1^\mu + \left(1-2(n\cdot u_1)^2\right) n^\mu}{\sqrt{1+(n\cdot u_1)^2}\sqrt{8(n\cdot u_1)^2-1}}\;.
 \eeq  

\begin{figure}
 \begin{center}
\includegraphics[width=4cm]{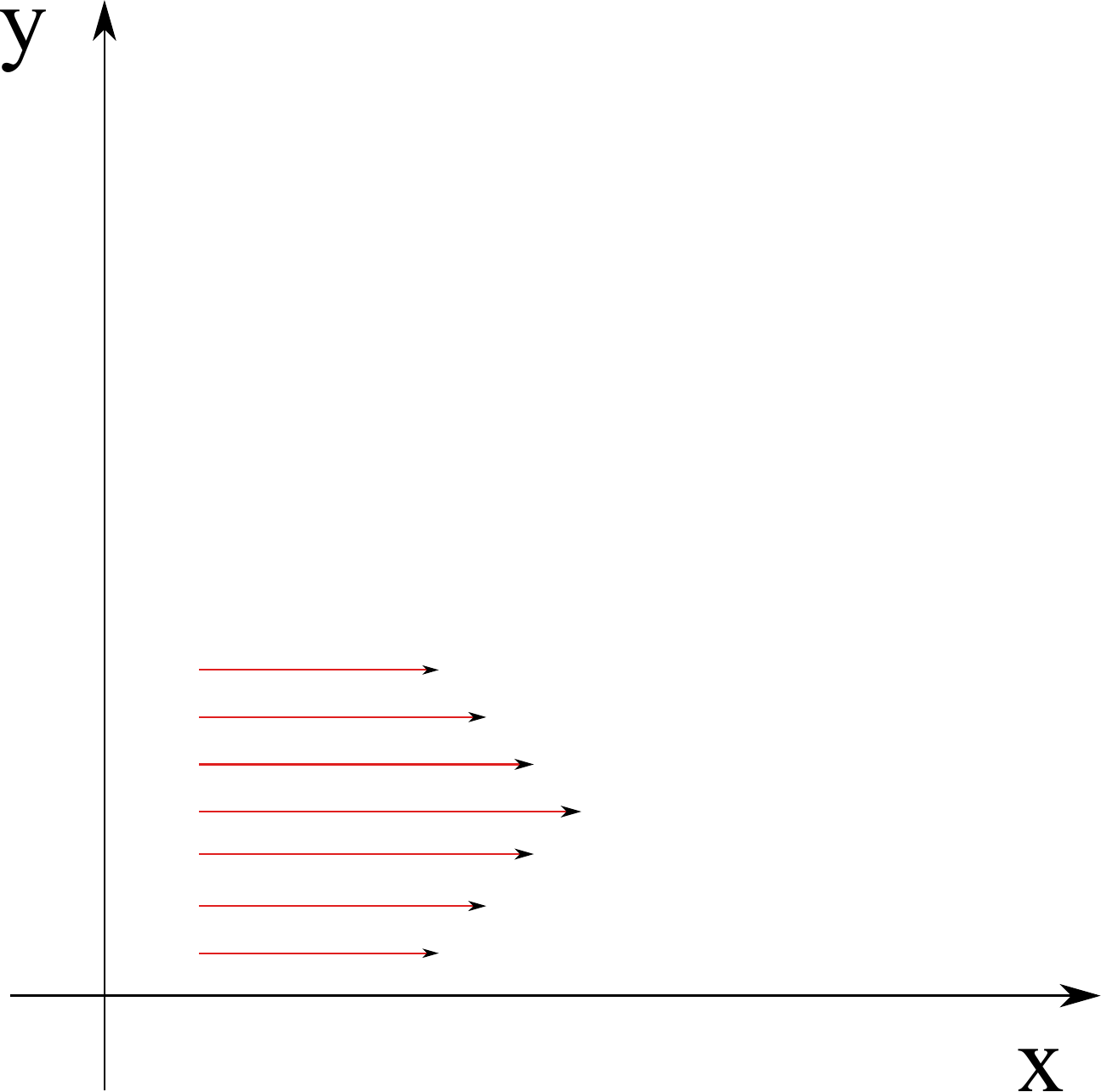}
 \end{center}
 \caption[]{A sketch of the shear or entropy  wave. }
\label{shearFig}
 \end{figure}
 
\subsection{Linearized analysis}

 Consider a flat stationary shock wave perpendicular to direction $x$, namely $n_\mu=(0,1,0,0)$. If a perturbation that deforms the shock wave is produced,
 it generates waves that move at most at the speed of sound  with respect to the fluid and therefore do not propagate into the supersonic region. So, at the linear level we need to solve
 \beqa
 \partial_\mu \delta T^{\mu\nu}_2 =0 , &\ \ \ &\mbox{subsonic side}, \\
 \delta n_\nu \left(T_1^{\mu\nu} - T_2^{\mu\nu}\right) =n_\mu\delta T_2^{\mu\nu} , &\ \ \ &\mbox{interface}.
 \eeqa
If we make a ripple on the surface given by\footnote{Of course the coordinates are real, we do the standard trick 
of defining them complex and taking the real part at the end.} $x=\zeta = \zeta_0 e^{i k_y y-i \omega t}$ then, to the first order in $\zeta$, the variation in the normal is
\beq
\delta n^\mu =-i \zeta (\omega,0,k_y ,0)\;.
\eeq
We also propose that
\beq 
\delta T^{\mu\nu}_2 = \delta \bar{T}^{\mu\nu}_2 e^{i(k_y y -\omega t+k_x x)}\;,
\eeq
with
\beq
k_\mu \delta \bar{T}^{\mu\nu}_2 = 0\;.
\eeq
There are two types of waves that solve these equations: shear waves and sound waves given by
\eqn{ShearSound}{
 \delta T^{\mbox{sh}} = 0 \;,\quad 
  \delta v_x^{\mbox{sh}}=-\frac{v_2 k_y}{\omega} \delta v_y^{\mbox{sh}}\;,\quad  k_x^{\mbox{sh}} = \frac{\omega}{v_2}\;, \cr
 \delta v_x^{\mbox{s}} = \frac{\delta T^{\mbox{s}}}{T_2} (1-v_2^2) \frac{k_x-\omega v_2}{\omega - k_x v_2} \;, \quad
 \delta v_y^{\mbox{s}} = \frac{\delta T^{\mbox{s}}}{T_2} (1-v_2^2) \frac{k_y}{\omega - k_x v_2} \;, \quad \cr k_x^2+k_y^2-\omega^2 = \frac{2}{1-v_2^2}(\omega-k_x v_2)^2 \;.
}
These are shear and sound waves boosted with velocity $v_2$. In a fluid at rest a shear wave is any time independent perturbation of the form
$\delta \vec{v} e^{i\vec{k}\vec{x}}$ with $\vec{k}\cdot \vec{x}=0$, which in the absence of viscosity is a solution of the equations of motion. 
Notice also that $\omega$ and $k_y$ are the same for both waves but the values of $k_x$ can be and actually are different. 
Finally matching at the interface we get the equations
\beqa
\delta v_x^{\mbox{sh}} + \delta v_x^{\mbox{s}} &=& -\frac{2}{3}i\zeta_0 \omega (1+3v_2^2) \;,\\
\delta v_y^{\mbox{sh}} + \delta v_y^{\mbox{s}} &=& i k_y \zeta_0 \frac{1-3v_2^2}{3v_2} \;,\\
\frac{\delta T^{\mbox{s}}}{T_2} &=& \frac{4}{3} \frac{i\zeta_0\omega v_2}{1-v_2^2}\;.
\eeqa
 Therefore we have six equations for six amplitudes 
($\delta v_x^{\mbox{sh}}$,$\delta v_y^{\mbox{sh}}$,$\delta v_x^{\mbox{s}}$,$\delta v_y^{\mbox{s}}$,$\delta T^{\mbox{s}}$,$\zeta_0$). 
The characteristic polynomial has to vanish, which gives a relation between $\omega$ and $k_y$:
\beq
\omega = \frac{1}{\sqrt{2}} \sqrt{\frac{1\mp v_2}{1\pm v_2}} k_y\;,
\eeq
plus another branch where $\omega$ has the opposite sign compared to $k_y$.  
Furthermore the dispersion relations determine:
\beq
k_x = -\frac{4v_2}{1-3v_2^2}\omega \pm \omega\;,
\eeq
for the sound wave and
\beq
k_x^{\mbox{sh}} = \frac{\omega}{v_2}\;,
\eeq
for the shear wave. What is quite interesting is that the sound that is produced has just two possible values of the ratio $\frac{k_x}{k_y}$
meaning that the direction of the outgoing sound is fixed.
It is also important to notice that $\omega$ is real meaning that there is no instability of the shock. In the ideal fluid approximation the 
oscillations of the shock are not damped and continuously produce sound taking energy from the fluid. It would be interesting to extend
this to the next order where viscosity is included and see how the oscillations are damped. 

\section{Discussion}\label{DISCUSSION}

In this paper we have continued our study of shock waves in strongly coupled plasmas.  Macroscopically, shocks are associated with a discontinuity of the temperature and pressure along some surface.  They typically arise in supersonic flows.  The appearance of the discontinuity signals a breakdown of the hydrodynamic approximation except for the case of weak shocks where viscosity is sufficient to resolve the discontinuity.  

Using AdS/CFT we have studied the asymptotic exponentially decaying tails of both strong and weak shocks for strongly coupled conformal plasmas in arbitrary dimensions.  In the rest frame of the shock front the asymptotic behavior is encoded in the penetration length as a function of the velocity $\ell = \frac{1}{q(v)}$.  We have numerically determined this curve for different dimensions and confirmed the analytic result that asymptotically, for velocities close to the speed of light, the penetration length scales as $\ell \sim \g ^{-2/d}$.  This means that the penetration length is an ultraviolet property of the medium. This is seen quite explicitly on the gravity side of the correspondence, 
where at large $\gamma$ the
spurious pole that controls the scaling approaches the boundary of AdS.  It would be interesting to examine other theories with gravity duals to find the value of the scaling exponent there.  In the context of QCD, our result suggests that a perturbative calculation may be enough to 
obtain the penetration length of the sound mode for ultrarelativistic shocks. 

For weak shocks, we have also examined the main properties of the analytic solution for the bulk 
black-hole metric
originally found in \cite{Khlebnikov:2010yt}.  We have established that there are no non-trivial Killing vectors in this background and therefore conventional surface gravity cannot be defined.  
Even the less strict condition of a non-expanding horizon does not apply to our solution.
Indeed, we have found that, to the leading order in the gradient expansion, the expansion of 
the horizon is directly proportional to the entropy production by a shock in the dual fluid.
Another interesting aspect of shock waves is generation of sound. For ideal fluid, the corrugation instability calculation shows that the sound is emitted at a definite angle to the direction in which
the shock propagates.

There are many interesting aspects of shocks that we have not addressed here.  The question whether a full gravity solution, analytical or numerical, exists has not been completely answered.  Such a solution would resolve the important question of what happens in the region where the linear perturbations diverge.  These solutions should also be interesting objects in gravity as we expect that they do not have constant surface gravity at the horizon.  Another direction of study would be to understand what happens to the corrugation (in)stability when viscosity is taken into account.  The inclusion of viscosity introduces a new length scale, the resolution length of the shock, and this means that multiple scale analysis has to be used.  The physical expectation is that the sound emitted by the shock will be damped by the viscous terms. 

 The recent calculation of thermodynamic functions and transport coefficients for superfluids in AdS/CFT, for example in \cite{Sonner:2010yx},\cite{Herzog:2011ec},\cite{Bhattacharya:2011ee},  also raises the question of existence of shocks in superfluids.  Shock waves in superfluids have been studied for example in \cite{Khalatnikov:1965}.  Two intriguing possibilities arise.  First, it is possible to have a discontinuity in the superfluid component of the velocity of the fluid while the normal component remains continuous.  Second, since the two sides of the shock have different temperatures it is possible to have different phases on the two sides of the shock.  We leave these interesting possibilities for future research.

\section*{Acknowledgements}\label{ACKN}

We acknowledge discussions with G.\; Horowitz and A.\; Ashtekar. This work was supported in part by the DOE under grant 
DE-FG02-91ER40681. The work of M.K. was also supported in part by the Alfred P. Sloan Foundation and by NSF under grant 
PHY-0805948 and a CAREER Award PHY-0952630.

\clearpage

\appendix

\section{Derivation of the equations of motion}\label{DERIVATION}

In this appendix we describe the method to derive the equations of motion for the sound channel.  The background metric is given by 
\eqn{BACKGRMET}{
ds^2 = \frac{1}{z^2} \h_{\m\n} dx^{\m}dx^{\n} +\frac{z^{2}}{z_{h}^{4}}\left(dt \frac{1}{\sqrt{1-v^2}}- dx \frac{v}{\sqrt{1-v^2}}\right)^2 + \frac{dz^2}{z^2(1-\frac{z^4}{z_{h}^4}) }\;,
} 
and the perturbation by 
\eqn{BHBOOSTPERTUR}{
ds^2 = \frac{e^{iqx}}{z^2} \left(h_{00} dt^2 +H_{11} dx^2 +2 H_{01} dt dx +H(d\vec{x}^2_{d-2})\right)\;.
}
Einstein's equations 
\eqn{GOTEINSTEIN}{
\tilde{G}_{MN} = R_{MN}+ d g_{MN} = 0\;,
}
give seven independent equations namely the $zz,zt,zx,tt,xx,xt,x_{d-2}x_{d-2}$ components of \eno{GOTEINSTEIN}.  We form the linear combination 
\eqn{GOTCURLL}{
\mathcal{L} = A^{MN}\tilde{G}_{MN}\;,
}
and we can choose four components of A, $A_{zz},A_{zx},A_{xx},A_{x_{d-2} x_{d-2}}$ to eliminate $H_{01},H_{11}$ and their derivatives.  The remaining two constants (the third is related to an overall normalization and is not relevant) are not sufficient to eliminate either of the remaining functions.  However a change to
\eqn{HOOTOZ}{
H_{00}(z) = Z(z) -g(z) H(z)\;,
}
with 
\eqn{GOTGZ}{
g(z) = 1+\frac{d-2}{2} \frac{(z/z_{h})^d}{1-v^2}\;,
}
allows one to eliminate $H(z)$ and write a decoupled equation for $Z(z)$ \eno{EOMZ}.  Then the other perturbations are determined from Einstein's equations where $Z(z)$ generically appears as a source term for the remaining fluctuations.  We do not give here the equations satisfied by the other perturbations but an interested reader can find the relevant equations for $d=4$ in appendix A of \cite{Khlebnikov:2010yt}.

\section{Regularity at the horizon}\label{REGULAR}

The boundary conditions imposed for the perturbations on the the horizon \eno{HORIZONBEH} are such that the perturbations diverge for positive imaginary values of $q$.  This is often the case for perturbations around a black hole.  In this appendix we wish to examine whether this divergence is an artifact of the coordinate system we use or an effect with a physical origin.  Let us restrict ourselves without loss of generality to the case of four dimensional plasmas and five dimensional black holes that was originally examined in \cite{Khlebnikov:2010yt}.  One can construct several covariant quantities that give information about the geometry. Ideally we would like to know all the eigenvalues of $R^{\a}_{\;\;\b \g \d}$.  However this goes beyond the scope of this appendix.  Out of the 25 invariants that can be built from the eigenvalues of the Riemann tensor in five dimensions we examine only one.   Since both $R_{\m\n}R^{\m\n}$ and $R$ are regular from the equations of motion we will restrict ourselves to calculating $R^{\a}_{\;\;\b \g \d}R_{\a}^{\;\;\b \g \d}$ which will give us  a first insight into regularity.  Evaluating  $R^{\a}_{\;\;\b \g \d}R_{\a}^{\;\;\b \g \d}$ to the first order in the perturbations \eno{PERTURBBH} and using the equations of motion which can be found in appendix (B) of \cite{Khlebnikov:2010yt} we find that 
\eqn{GOTRSQUARE}{
R^{\a}_{\;\;\b \g \d}R_{\a}^{\;\;\b \g \d} =& \frac{8\left( 5z_{h}^8 + 9z^8 \right)}{z_{h}^8} -288 \frac{z^8}{z_{h}^8} e^{-iqx} H(r) +288\frac{(1-v^2)(z_{h}^4-z^4)z^5}{\left(z^4-3z_{h}^4(1-v^2)\right)^2}e^{-iqx} Z'(z)- \cr 
  &-\left( 288(1-v^2) \frac{z^8 \left(3z^4+z_{h}^4 (3v^2-5)\right)}{z_{h}^4\left(z^4-3z_{h}^4(1-v^2)\right)^2}+48 q^2 \frac{z^6}{z_{h}^8\left(z^{4}-3z_{h}^4(1-v^2)\right)}\right)e^{-iqx}Z(z)\;.
}
The expression \eno{GOTRSQUARE} has been evaluated on shell, meaning that all derivatives higher than the second have been substituted using Einstein's equations.  We now have to use the behavior of the perturbations close to the horizon to examine whether the invariant quantity diverges.  Using table from \cite{Khlebnikov:2010yt}, which for reference is reproduced here in table \eno{TABLEWAVE}  we find that close to the horizon 
\eqn{GOTRSQUAREHOR}{
R^{\a}_{\;\;\b \g \d}R_{\a}^{\;\;\b \g \d} \sim e^{iqx} \left(z-z_{h} \right)^{\frac{iq v x}{4\sqrt{1-v^2}}}   \;.
}
Thus, there is a curvature singularity at the horizon.  A way out of this would be to suppose that as the metric diverges close to the horizon the non linearities become important and smooth out the curvature.  Such a line of thought can only be pursued numerically and we leave it for future work.

The result of this appendix may appear to be in contradiction to the construction of a smooth metric describing weak shock waves in \cite{Khlebnikov:2010yt}.  However this is not true.  
The reason is that, in the  Eddington-Finkelstein coordinates used in \cite{Khlebnikov:2010yt}, 
the location of the potential divergence at the horizon is pushed to the past infinity. To see
that, let us use the coordinate change from the Poincare coordinates to the Eddington-Finkelstein ones close to the horizon:
\eqn{COORDXCHANGE}{
x = \tilde{x} - \frac{v}{4\sqrt{1-v^2}} \log(z_{h}-z)\;.
}   

\begin{figure}
 \begin{center}
\includegraphics[width=6cm]{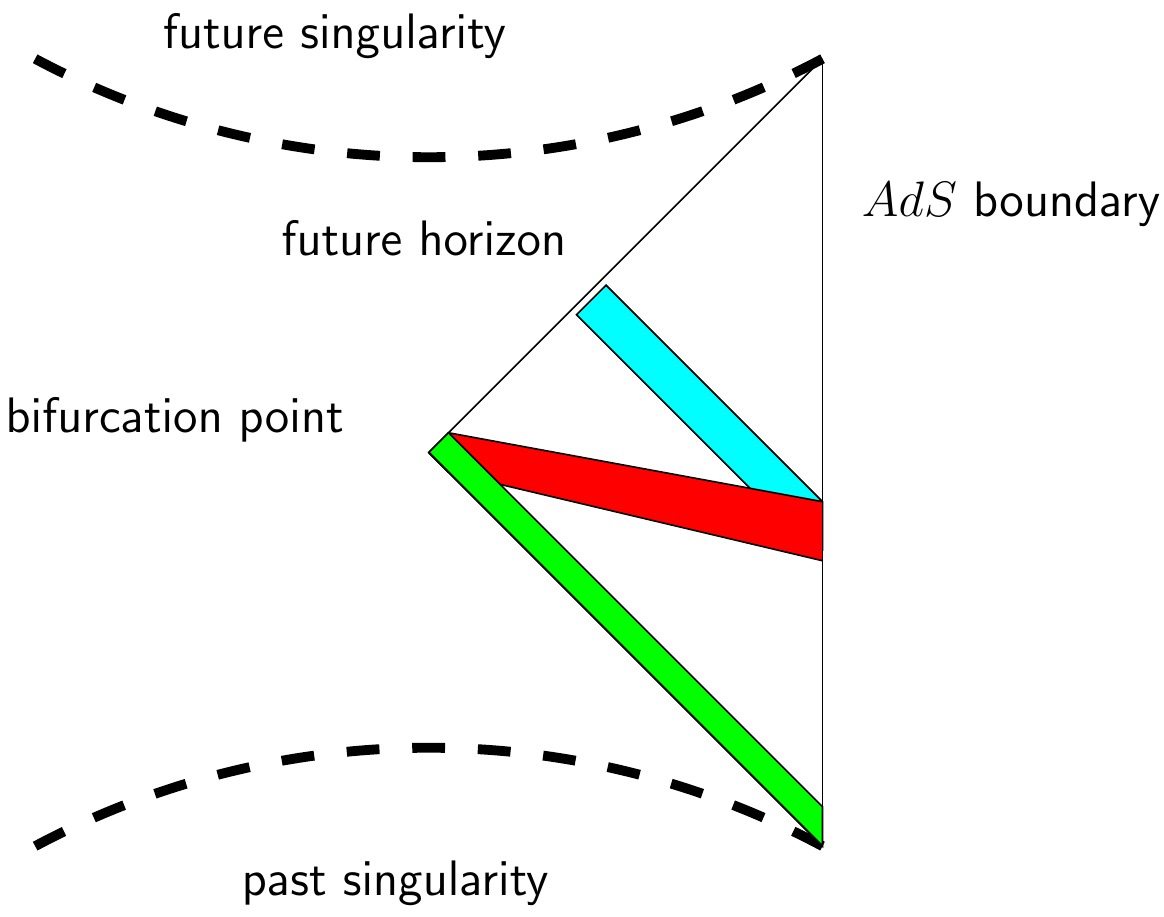}
 \end{center}
 \caption[]{A rough sketch of the Penrose diagram for a boosted black hole.  Only the relevant parts are shown.  Starting from a patch in the boundary and keeping $x,t$ constant in Eddington-Finkelstein coordinates we construct tube shown in cyan.  On the other hand if we keep $x,t$ constant in Poincare or Schwarzschild coordinates we obtain the tube shown in red.  The potential divergences arise in the region close to the bifurcation point.  In order to explore the same problematic region in Eddington-Finkelstein coordinates one has to take a spatial and temporal infinity limit .  We believe that the resolution of the divergences will appear only in the full non-linear solution.}
\label{BHole}
 \end{figure}

This changes the boundary conditions at the horizon  from \eno{HORIZONBEH}  to 
\eqn{BOUNDHEF}{
Z \sim \mbox{const}. \;, ~~~Z\sim (z_{h}-z)^{\frac{-iqv}{2\sqrt{1-v^2}}} \;.
}
That is the infalling condition becomes a constant and the outgoing becomes a non-analytic function for generic values of $q$.  This has been used in \cite{Khlebnikov:2010yt} to argue that the boundary value problem set by \eno{EOMZ},\eno{HORIZONBEH} is well posed.  It should be noted here that the location of horizon in Poincare coordinates $z=z_{h}$ with $x,t$ finite translates into $z=z_{h}$ with $\tilde{x},\tilde{t}$ going to $-\infty$.  That is there is no contradiction with the construction in the Eddington-Finkelstein coordinates since we are not examining the same geometric region.  A schematic of this argument can be found in figure \eno{BHole}.
  
Having calculated the invariant \eno{GOTRSQUARE}, we can also examine its behavior close to the pole of the sound propagation equation \eno{EOMZ}. From the analytical solution close to the pole, we know that it is a regular point. So, we do not expect any divergence, even though some of the 
coefficients in \eno{GOTRSQUARE} have second or first-order poles at $z=z_{f}$. 
Using the solution around $z=z_{f}$ \eno{ZFUNSERIES},\eno{GOTC1},\eno{GOTC2} we find that 
the curvature invariant is indeed finite at this point.

\begin{table}[ht]\label{TABLEWAVE}
\begin{center}
\begin{tabular}{||c || c | c | c|}
\hline
   & $r\rightarrow \infty $&$r\rightarrow r_h$& $r\rightarrow r_{f}$  \\
\hline
 $Z(r)$  & $r^{-4}$ & $(r-r_{h})^{-i \frac{qv}{4r_{h} \sqrt{(1-v^2)}}} $ & $\z_{0} + \z_{1}(r-r_{f}) +\z_{2}(r-r_{f})^2+\cdots$ \\
\hline
  $H_{00}(r)$ & $r^{-4}$ &$\frac{2v^2}{3v^2-2} (r-r_{h})^{-i \frac{qv}{4r_{h} \sqrt{(1-v^2)}}}$, $v< \sqrt{\frac{2}{3}}$  &$h^{(0)}_{0}+h^{(0)}_{1}(r-r_{f})+\cdots$\\
\hline
  $H(r)$ & $r^{-4}$ & $\frac{1-v^2}{3v^2-2}(r-r_{h})^{-i \frac{qv}{4r_{h} \sqrt{(1-v^2)}}} $, $v<\sqrt{\frac{2}{3}}$ &$h_{0}+h_{1}(r-r_{f})+\cdots$\\
               &                &    $\tilde{C}_{1}  \sqrt{r-r_h} $, $v\ge \sqrt{\frac{2}{3}} $                                                                         &  \\
\hline
  $\tilde{H}(r)$ & $r^{-9}$ &  $-\frac{2iv(-iqv+2(1-2v^2)\sqrt{1-v^2}-q^2(1-v^2)^{3/2})}{(3v^2-2)(qv-4i\sqrt{1-v^2})}\cdot$&$\tilde{h}_{0} +\tilde{h}_{1}(r-r_f)+\cdots$ \\
           & & $\cdot(r-r_{h})^{-i \frac{qv}{4r_{h} \sqrt{(1-v^2)}}}$,  $v<\sqrt{\frac{2}{3}}$  & \\
           & &  $\tilde{C}_{2} \sqrt{r-r_{h}}$ ,$v\ge \sqrt{\frac{2}{3}}$                   &\\
\hline
  $H_{01}(r)$ & $r^{-9}$ &  &$h^{(01)}_{0}+h^{(01)}_{1}(r-r_{f})+\cdots $ \\
\hline
  $H_{11}(r)$ & $r^{-9}$ &  &$h^{(11)}_{0}+h^{(11)}_{1}(r-r_{f})+\cdots $\\
\hline
\end{tabular}
\caption{The behavior of the perturbations close to the boundary, the horizon and the pole of the equation \eno{EOMZ} is given.  To connect with the notation used in the present paper, we have $r=\frac{1}{z}$, $r_{h}=\frac{1}{z_{h}}$.  Here, $r_{f}=\frac{1}{z_{f}}=\frac{1}{3^{1/4}}\frac{r_{h}}{(1-v^2)^{1/4}}$ is the inverse of the location of the pole of \eno{EOMZ}.  The boundary conditions for the perturbations at the boundary $r\rightarrow \infty$ are that the metric is unchanged from the Minkowski metric. At the horizon the condition is that the asymptotic behavior corresponds to an infalling graviton in the unboosted black hole metric. In the linearized approximation the overall scaling factor does not appear and we only show the power law behavior. The normalization of the perturbations close to the horizon are relative to the normalization of $Z(r)$.  The normalization of $Z(r)$ is taken to be $1$ for the factor multiplying $(r-r_{h})^{-i \frac{qv}{4r_{h} \sqrt{(1-v^2)}}} $. For $v\ge \sqrt{\frac{2}{3}}$ the relative coefficients $\tilde{C}_{1,2}$ are not computed.    }
\end{center}
\end{table}

\clearpage

\bibliographystyle{xbib}
\bibliography{shock2}
\end{document}